\documentclass[aps,prb,floatfix,amsmath,amssymb,preprint,eqsecnum,nofootinbib]{revtex4}
\usepackage{graphicx}% Include figure files
\usepackage{dcolumn}% Align table columns on decimal point
\usepackage{bm}% bold math
\usepackage{setspace}   % controllabel line spacing
\usepackage{epstopdf}
\usepackage{bbm}
\newcommand{\beq}{\begin{equation}}
\newcommand{\eeq}{\end{equation}}
\newcommand{\ba}{\begin{array}{ccc}}
\newcommand{\ea}{\end{array}}
\newcommand{\nn}{\nonumber}
 \renewcommand{\d}{\partial}
\def\bea{\begin{eqnarray}}
\def\eea{\end{eqnarray}}

\def\<{\langle}
\def\>{\rangle}
\usepackage{amsmath}
\usepackage{amssymb}
\begin{document}
\title{Valence bond solid order near impurities in two-dimensional quantum antiferromagnets}

\author{Max~A.~Metlitski}
\email{mmetlits@fas.harvard.edu} \affiliation{Department of Physics,
Harvard University, Cambridge MA 02138, USA}

\author{Subir Sachdev}
\email{subir_sachdev@harvard.edu} \affiliation{Department of
Physics, Harvard University, Cambridge MA 02138, USA}

\date{\today\\[24pt]}

\begin{abstract}
Recent scanning tunnelling microscopy (STM) experiments on underdoped cuprates have  
displayed modulations in the local electronic
density of states which are centered on a Cu-O-Cu bond (Kohsaka {\em  
et al.}, Science {\bf 315}, 1380 (2007)).
As a paradigm of the pinning of such bond-centered ordering in  
strongly correlated systems, we present
the theory of valence bond solid (VBS) correlations near a single  
impurity in a square lattice antiferromagnet.
The antiferromagnet is assumed to be in the vicinity of a quantum  
transition from a magnetically
ordered N\'eel state to a spin-gap state with long-range VBS order.
We identify two distinct classes of impurities: ({\em i\/}) local  
modulation in the exchange constants,
and ({\em ii\/}) a missing or additional spin, for which the impurity  
perturbation is represented
by an uncompensated Berry phase. The ``boundary" critical theory for  
these classes is developed: in the second
class we find a ``VBS pinwheel" around the impurity,  
accompanied by a suppression in the VBS susceptibility.
Implications for numerical studies of quantum antiferromagnets and  
for STM experiments on the cuprates are noted.
\end{abstract}

\maketitle

\section{Introduction}

A number of recent scanning tunnelling microscopy experiments have highlighted spatial modulations in the local density of states in the cuprate
compounds, nucleated by external perturbations. In Ref.~\onlinecite{versh}, the spatial modulation was 
observed in the normal state above $T_c$, presumably nucleated by impurities. In Refs.~\onlinecite{hoffman,fischer1,fischer2}, 
the order was found in a halo around vortices, which were in turn pinned by impurities. Most recently, in Ref.~\onlinecite{kohsaka}, 
similar charge-ordering
patterns were found to be ubiquitous in the underdoped cuprates at low temperatures, and it was established that
the charge ordering was ``bond-centered", and had an anisotropic structure similar to a valence bond solid state \cite{adrian,kivelson,poilblanc,vojta}.

In the light of these observations, it is of general interest to study the appearance of varieties
of charge order (including ``valence bond solid" (VBS) order \cite{poilblanc,vojta,vbscharge}) near impurities in strongly correlated systems. 
For superfluid states, such a theory has been
presented in earlier work \cite{balents1,balents2}, and compared quantitatively with some of the above experiments. 
It was argued that the charge order was linked to quantum fluctuations of vortices/anti-vortices in the superfluid order.
Consequently, the problem mapped onto the pinning of the vortices by impurities, and the quantum zero-point motion of vortices
about the pinning site. In both zero and non-zero magnetic fields, enhanced charge order was found in the spatial region
over which the vortex executed its zero-point motion \cite{balents1}. This charge order was present even when the net vorticity was
zero everywhere (as is the case in zero magnetic field): the vorticity cancelled between the vortex and anti-vortex fluctuations,
but the charge order did not.

\begin{figure}[t]
\centering \includegraphics[width=3in]{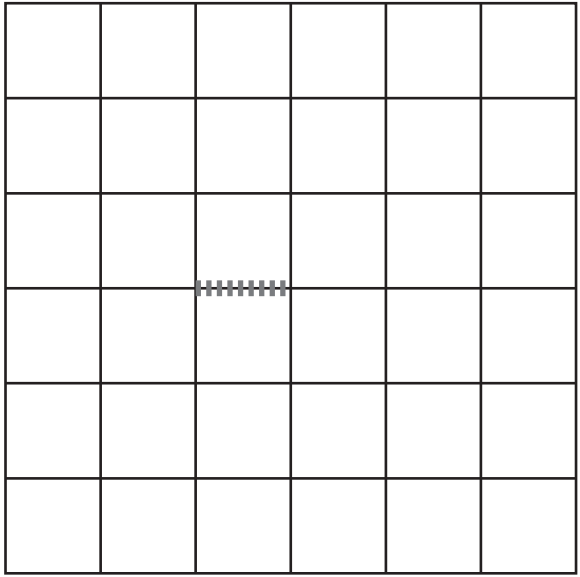}
\caption{A modulated exchange impurity which is described in Section~\ref{sec:exch}. The dashed line indicates a different value of
the antiferromagnetic exchange constant. We expect VBS order to be enhanced near such an impurity, because the modulated
exchange will lock in a preferred orientation and offset of the VBS state.}
\label{fig:bond}
\end{figure}

\begin{figure}[t]
\centering \includegraphics[width=6in]{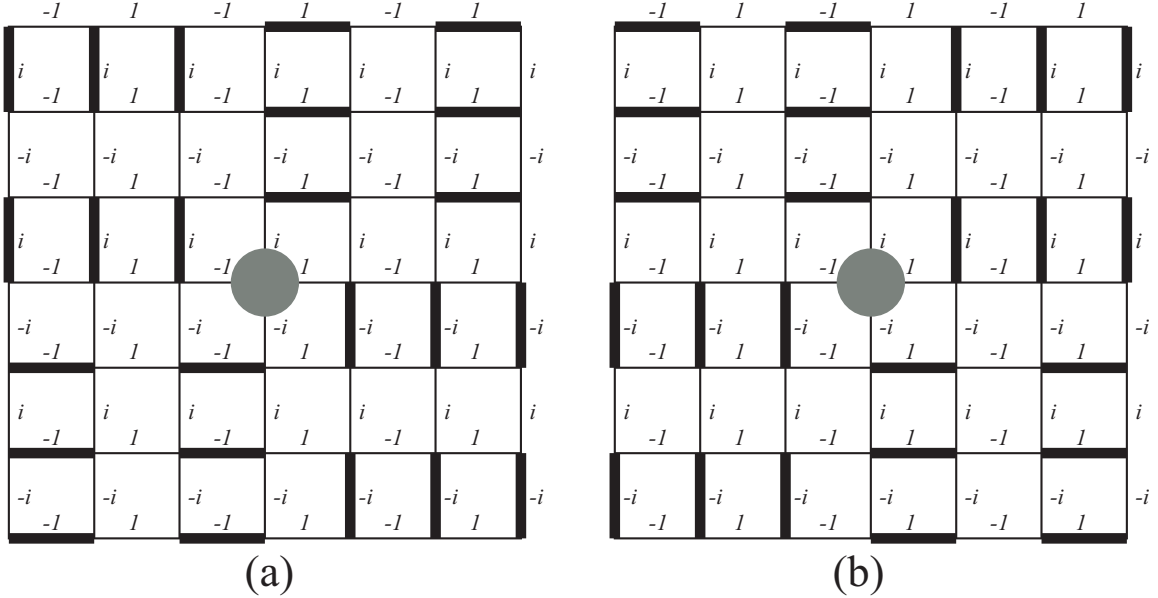}
\caption{A vacancy (the shaded circle) 
in a square lattice quantum antiferromagnet which is described in Section~\ref{sec:berry} and the remainder of the paper. 
The thick lines represent
singlet bonds between the spins. The local value of the VBS order is measured by the phase factors on the singlet bonds. Moving anti-clockwise
from the right in both figures, we observe that VBS order cycles as $1 \rightarrow i \rightarrow -1 \rightarrow -i$ in (a),
and as $i \rightarrow -1 \rightarrow -i \rightarrow 1$ (b). Thus, {\em both\/} configurations are ``vortices" in the VBS order, which we name ``VBS pinwheels" (these VBS pinwheels
are ``dual" to the vortices in the superfluid/N\'eel order that are discussed in the beginning of the paper). 
Anti-pinwheels in the VBS
order appear only
around vacancies on the other sublattice; in other words, VBS pinwheels transform to VBS anti-pinwheels under translation
by a single site---see Fig~\ref{fig:antivortex} later.}
\label{fig:vortex}
\end{figure}

This paper will present an extensive field-theoretic analysis of a paradigm of the problem of charge order near impurities
in correlated systems. We will consider insulating $S=1/2$ antiferromagnets on the square lattice, across a quantum 
phase transition from the magnetically ordered N\'eel state, to a spin-gap valence bond solid (VBS) state \cite{rs1,rs2,senthil1,senthil2}. 
By representing the $S=1/2$
spins as hard-core bosons, our results can be reinterpreted as applying to the superfluid-insulator transition of bosons at half-filling
on the square lattice: the N\'eel state of the antiferromagnet maps onto the superfluid state of the bosons, while the VBS state
maps onto a Bose insulator with bond-centered charge order. The bond-centered charge correlations in the underdoped
cuprates  now appear to have two possible physical mechanisms (``disordered" antiferromagnet/superfluid), but 
it was argued in Ref.~\onlinecite{balents3} that they represent the same underlying physics. Our results here will go beyond the earlier work \cite{balents1,balents2} 
in two important respects:\\
({\em i\/}) We will describe the critical singularities in the impurity-induced VBS/charge order at the quantum critical point, and\\
({\em ii\/}) We will consider a wider class of impurity perturbations. In the previous work \cite{balents1,balents2}, 
an ``impurity" was assumed to be a generic deformation of the underlying Hamiltonian which broke its space group symmetry. For the N\'eel-VBS transition, such an impurity is realized {\em e.g.\/} by the modulation in the magnitude of a particular exchange coupling -- see Fig~\ref{fig:bond}. 
We briefly will discuss the critical singularities describing the {\em enhancement\/}  of VBS order near such an impurity in Section~\ref{sec:exch} below; these results have a natural extension to the models
of charge order near the superfluid-insulator transition discussed above. However, the primary focus of the present paper is on a distinct class of impurities, in which the valence-bond structure of the non-magnetic ground state of the antiferromagnet 
is more strongly disrupted, and a ``Berry phase" contribution
of an unpaired spin is the crucial impurity-induced perturbation \cite{science,kolezhuk}. Such impurities are realized by replacing the $S=1/2$ Cu spins in 
antiferromagnets by a non-magnetic Zn ion, or a $S=1$ Ni ion (see Fig.~\ref{fig:vortex}). For the superfluid-insulator transition, such an impurity is a site
from which particles are excluded, and so a local ``phase-shift" is induced in the charge order of the insulating state  (replacing a Cu atom by Zn or Ni is expected to have the desired ``Cooper pair" exclusion effect\cite{balents3}). Our main results
will include a description of the {\em suppression\/} of VBS order near such ``Berry phase" impurities: these results are summarized in 
Section~\ref{sec:berry} below, and described in the body of the paper. 
A simple sketch of how such an impurity disrupts the VBS order is shown in Fig.~\ref{fig:vortex}; this figure builds upon the dual theory
of spinons in the VBS state developed by
Levin and Senthil \cite{levinsenthil}. The bulk of this paper will
describe how quantum fluctuations of the type sketched in Fig.~\ref{fig:vortex} lead to a modification of the scaling dimension
of the VBS order in the vicinity of the vacancy.

The remainder of the paper will be presented in the language of the N\'eel-VBS transition in quantum antiferromagnets,
For this model, a field theoretic description of the vicinity of the quantum critical point \cite{senthil1,senthil2,SachdevMurthy,mv} is provided by the $\mathbb{CP}^{N-1}$ theory at $N=2$:
\begin{equation}
\mathcal{S} = \int d^2 x d \tau \left[ |(\partial_\mu - i A_\mu)z_\alpha |^2 + s |z_\alpha|^2 + \frac{g}{2} \left( |z_\alpha|^2 \right)^2 + \frac{1}{2e^2} (\epsilon_{\mu\nu\lambda} \partial_\nu A_\lambda)^2 \right]. \label{cpn}
\end{equation}
Here $\mu,\nu\,\lambda$ are spacetime indices, 
$z_\alpha$, $\alpha = 1 \ldots N$ is a complex scalar which is a SU($N$) fundamental, and $A_\mu$ is a non-compact U(1) gauge field.
The N\'eel order of the antiferromagnet is $n^a = z^\dagger T^a z$, where $T^a$ is a SU($N$) generator. The SU($N$) symmetry is
spontaneously broken in the N\'eel phase, $\langle n^a \rangle \neq 0$, which is realized for $s<s_c$, where $s_c$ is the critical
value of the tuning parameter, $s$, for the quantum phase transition. For $s>s_c$, the $\mathbb{CP}^{N-1}$ theory above describes a U(1)
spin liquid state of the antiferromagnet, with gapped spinons $z_\alpha$ and a gapless, U(1) photon. However, as has been argued at
length elsewhere \cite{rs1,rs2}, lattice effects not included in the continuum field theory (\ref{cpn}) eventually render the U(1) spin liquid
unstable to spinon confinement and fully gapped state with VBS order. The VBS order parameter, $V$, is an operator \cite{rs1,senthil2} which creates a {\em Dirac monopole\/} with total flux $2\pi$ in the U(1) gauge field $A_\mu$. This paper will therefore be concerned with correlations
of the monopole/VBS operator $V$ under the field theory $\mathcal{S}$ after including the impurity perturbations described below.
The bulk scaling dimension of the monopole operator at the $s=s_c$ critical point will make frequent appearances in our analysis, and so we define this as
\begin{equation}
\Delta^V = \mbox{dim}[V(\vec{x},\tau)]~~\mbox{in the theory $\mathcal{S}$ without an impurity.}
\label{dimv}
\end{equation}

The following subsections will now describe the two classes of impurity perturbations to the theory $\mathcal{S}$ shown in
Figs.~\ref{fig:bond} and~\ref{fig:vortex} respectively.

\subsection{Modulated exchange}
\label{sec:exch}

A modulation in the magnitude of an exchange constant in the underlying antiferromagnet (see Fig.~\ref{fig:bond}) breaks the lattice space group
symmetry, but preserves the spin rotation symmetry. Also, the number of spins on each sublattice is preserved, so no ``Berry phase"
term is expected. Consequently, we need to consider all local perturbations to $\mathcal{S}$ which preserve the required
symmetries. The simplest allowed possibility is a local shift in the position of the critical point. For an impurity at the 
spatial origin, $x=0$, this would lead to a term
\begin{equation}
\widetilde{s} \int d\tau |z_\alpha (\vec{x}=0,\tau)|^2
\label{dimz}
\end{equation}
However, a simple computation \cite{qimp1} shows that $\widetilde{s}$ is very likely an irrelevant perturbation at the bulk critical point.
We have $\mbox{dim}[\widetilde{s}] = 1 - (D - 1/\nu)$, where $D=3$ is the spacetime dimension, and $\nu$ is the correlation
length exponent of $\mathcal{S}$. Because it is almost certainly
the case that $\nu > 1/2$, we conclude that $\widetilde{s}$ is irrelevant. However, a more interesting perturbation is that considered 
in previous work \cite{balents1,balents2} on the superfluid-insulator transition. In the present context, this perturbation follows from the fact that with
broken space group symmetry, a linear coupling to the monopole operator is permitted. So we have the impurity action
\begin{equation}
\widetilde{\mathcal{S}}_{\rm imp} = \int d \tau \left[ h^\ast \, V(\vec{x} = 0, \tau) + \mbox{c.c.} \right]
\end{equation}
where $h$ is a complex-valued constant whose value depends upon the details of the modulated exchange near $x=0$. 
Now the renormalization group (RG) flow of $h$ follows from Eq.~(\ref{dimv}) to linear order
\begin{equation}
\frac{dh}{d\ell} = (1 - \Delta^V) h + \mathcal{O} (h^2)
\label{hrg}
\end{equation}
The remainder of this subsection will analyze the correlations of the monopole/VBS operator $V(\vec{x},\tau)$ in the theory
$\mathcal{S} + \widetilde{\mathcal{S}}_{\rm imp}$.

First, let us consider the likely possibility that $\Delta^V < 1$. In this case, $h$ is a relevant perturbation, and higher order corrections
to Eq.~(\ref{hrg}) cannot be ignored. By analogy with results in the theory of boundary critical phenomena \cite{diehl1}, and in particular with the theory of the ``extraordinary" transition \cite{bray,diehl2,diehl3}, we conclude that a likely possibility is that the RG flow is to strong coupling, to a fixed
point with $|h| = \infty$. In this, case some powerful statements on the correlations of $V(\vec{x}, \tau)$
can be immediately made. It is useful to express the correlations in the vicinity of the impurity by an operator product expansion (OPE). In general, this expansion will have the structure
\begin{equation}
\lim_{|\vec{x}| \rightarrow 0} V(\vec{x}, \tau) \sim  |\vec{x}|^{\Delta^V_{\rm imp}} \, V_{\rm imp} (\tau)
\label{vimp1}
\end{equation}
where $V_{\rm imp}$ is an operator localized on the impurity site, and $\Delta^V_{\rm imp}$ is the difference
in scaling dimensions between $V$ and $V_{\rm imp}$. Specifically, Eq.~(\ref{vimp1}) implies
\begin{equation}
\Delta^V = - \Delta^V_{\rm imp} + \mbox{dim}[V_{\rm imp}].
\label{vimp2}
\end{equation}
Now at a $|h| = \infty$ fixed point, we expect that fluctuations of $V$ near the impurity are strongly suppressed, and so it is a
reasonable conclusion that
$V_{\rm imp}$ is just the identity operator
\begin{equation}
V_{\rm imp} = \mathbbm{1}.
\end{equation}
Consequently, $\mbox{dim}[V_{\rm imp}]=0$, and we have our main result
\begin{equation}
\Delta^V_{\rm imp} = - \Delta^V.
\label{vimp3}
\end{equation}
The combination of Eq.~(\ref{vimp1}) and (\ref{vimp3}) appears to be a promising route to measuring
the scaling dimension of a monopole operator in numerical studies of quantum antiferromagnets. 

To complete our analysis of modulated exchange, we also address the case with $\Delta^V > 1$. In this situation,
by Eq.~(\ref{hrg}), the perturbation $h$ is irrelevant, and so we may compute the consequences of $h$ by perturbation
theory. Computing correlations to first order in $h$ we see that Eq.~(\ref{vimp1}) is now replaced by
\begin{equation}
\lim_{|\vec{x}| \rightarrow 0} V(\vec{x}, \tau) \sim  h \, |\vec{x}|^{-2\Delta^V + 1} 
\label{vimp4}
\end{equation}

\subsection{Missing spin}
\label{sec:berry}

Next we will consider the behavior of the
monopole/VBS operator $V$ near the missing spin impurity 
illustrated in Fig.~\ref{fig:vortex}. As discussed in some detail in Ref.~\onlinecite{kolezhuk}, the dominant
consequent of such an impurity is an exactly marginal perturbation to $\mathcal{S}$ given by
\begin{equation}
\mathcal{S}_{\rm imp} = i Q \int d\tau A_\tau (\vec{x} = 0, \tau) \label{simp}
\end{equation}
where $Q$ is a ``charge" characterizing the impurity. The value of $Q$ does not flow under the RG,
and so $Q$ is a pure number which controls all universal characteristics of the impurity response. For an impurity of Fig. \ref{fig:vortex} with a single missing spin, $Q = \pm 1$.
The remainder of this paper presents an analysis of the critical properties of the $\mathcal{S} + \mathcal{S}_{\rm imp}$
defined in Eqs.~(\ref{cpn}) and (\ref{simp}).

The magnetic correlations of the theory $\mathcal{S}+\mathcal{S}_{\rm imp}$ (and of a related theory \cite{lars}) have been computed in previous 
papers \cite{kolezhuk,MS} which obtained the scaling dimensions of the N\'eel order parameter, $n^a$, 
and of the uniform magnetization density in the vicinity of the impurity. It was found that the impurity significantly
enhanced the local magnetic susceptibilities. For the case of double-layer antiferromagnets, which have magnetic ordering transitions
described by Landau-Ginzburg-Wilson theory, such impurity magnetic correlations have also been computed by similar
methods \cite{science,qimp1,qimp2}, and found to be in excellent agreement with numerical studies \cite{hoglund1,hoglund2,hoglund3}.

\begin{figure}[t]
\centering \includegraphics[width=3.5in]{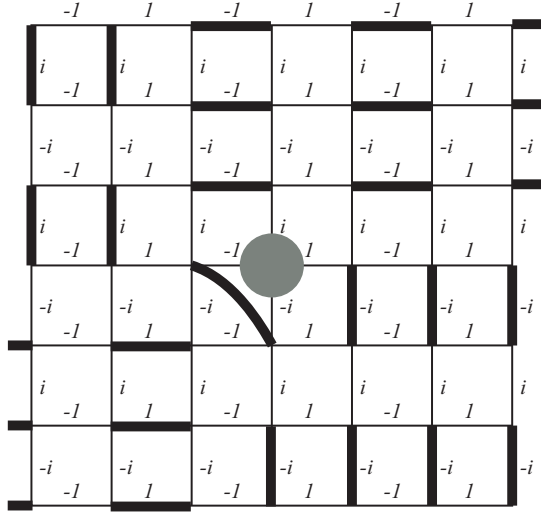}
\caption{A VBS {\em anti\/}-pinwheel in the presence of an impurity with a charge $Q$ opposite in sign to 
that required by Eq.~(\ref{vimp5}). This configuration has a higher energy cost than the VBS pinwheel configurations
in Fig.~\ref{fig:vortex}.}
\label{fig:antivortex}
\end{figure}

This paper will describe the ``charge-order" correlations of the theory $\mathcal{S}+\mathcal{S}_{\rm imp}$ by a computation
of the OPE of the monopole/VBS operator $V( \vec{x}, \tau)$ as $\vec{x} \rightarrow 0$. Our
principal result is that the OPE is modified from the form in Eq.~(\ref{vimp1}) to
\begin{equation}
\lim_{|\vec{x}| \rightarrow 0} V(\vec{x}, \tau) \sim  |\vec{x}|^{\Delta^V_{\rm imp}} \, e^{-i Q \theta} \, V_{\rm imp} (\tau)
\label{vimp5}
\end{equation}
where $\theta$ is the azimuthal angle of $\vec{x}$.
There are two important changes from Eq.~(\ref{vimp1}). The first is that $V_{\rm imp}$ is no longer a trivial unit operator,
but a fluctuating impurity degree of freedom with a non-trivial scaling dimension. The second is the presence of the $e^{-iQ \theta}$
factor, which indicates a $Q$-fold winding in the phase of the VBS order parameter around the impurity. The sketches in Fig.~\ref{fig:vortex}
give a simple physical interpretation of this winding in terms of the valence bond configurations of the underlying 
antiferromagnet: we will call this vortex-like winding in the VBS order a ``VBS pinwheel." Also, as we discussed earlier \cite{kolezhuk}, the sign of $Q$ is determined by the sublattice location of the missing spin.
Thus, the result Eq.~(\ref{vimp5})  indicates that VBS pinwheels will occur preferentially around impurities on one sublattice,
while VBS anti-pinwheels occur around impurities on the other sublattice. This same result is also obtained from the intuitive
microscopic pictures in Fig.~\ref{fig:vortex}. Also, we show in Fig.~\ref{fig:antivortex} an illustration of an anti-pinwheel 
in the presence of an impurity on the disfavored sublattice: the same sublattice bond indicates that this configuration has
a higher energy.

Apart from establishing the form of Eq.~(\ref{vimp5}), we will also describe computations of the exponent
$\Delta^V_{\rm imp}$. There are general reasons for  expecting that $\Delta^V_{\rm imp} > 0$, and this will be the
case in the explicit result we obtain. This positive value of $\Delta^V_{\rm imp}$ characterizes the {\em suppression\/}
of VBS order near the impurity, and should be contrasted with the negative value in Eq.~(\ref{vimp3}) for the impurity in Fig.~\ref{fig:bond}.

Our analysis will begin in Section~\ref{sec:CPN} by a large $N$ analysis of the theory $\mathcal{S}+\mathcal{S}_{\rm imp}$
with full SU($N$) spin symmetry. We will establish Eq.~(\ref{vimp5}) in this limit. We will also find that the $N=\infty$ limit
(at fixed $Q$) of the exponent $\Delta^V_{\rm imp}$ vanishes, but we will not evaluate the subleading correction in the $1/N$ expansion here.

The remainder of the paper will explore another approach to estimating $\Delta^V_{\rm imp}$. This relies \cite{senthil1,senthil2} 
on examining the ``easy-plane"
limit of the $\mathbb{CP}^{N-1}$ model, in which the global SU($N$) spin symmetry is reduced to $U(1)^{N-1}$. With this simplification
to an abelian global symmetry, an explicit duality transformation of the theory becomes possible. In the dual theory,
the monopole/VBS operator $V$ has a local expression in terms of the dual fields, and so this facilitates the analysis of the impurity
critical property. We will begin the dual analysis in Section~\ref{sec:dualityN1} 
by considering the simplest $N=1$ case \cite{sj}: this model describes the onset of VBS order in a
$S=1/2$ quantum antiferromagnet in the presence of a staggered magnetic field \cite{senthil2}, and 
is the simplest setting in which several technical issues can be described. 
We then extend the analysis to general $N$ in Section~\ref{sec:EasyPlane}. The exponent $\Delta^V_{\rm imp}$ will be estimated
in these sections by a self-consistent theory of Gaussian fluctuations about a mean-field state; in the physically interesting
case of $N=2$ and $Q=1$, which describes both the easy plane antiferromagnet and the boson superfluid/insulator transition, we obtain the estimate
\beq \Delta^{V}_{\mathrm{imp}} \approx 0.57, \quad N = 2, Q = 1\eeq
Our analysis of the easy plane theory in Section~\ref{sec:EasyPlane} also exhibits certain features which we do not
expect to be shared by the case with global SU($N$) symmetry: for $Q/N = 1/2$, we find VBS-vortex solutions in which the $e^{-iQ \theta}$
factor in Eq.~(\ref{vimp5}) is replaced by $e^{-i \ell \theta}$ with the integer $-Q \leq \ell \leq Q$. In the self-consistent
theory we present here, all the values of $\ell$ are degenerate, but we expect these degeneracies are partially lifted in
the full easy-plane theory. These issues are discussed further in Section~\ref{sec:EasyPlane} and in a forthcoming paper.

\section{$1/N$ Expansion of the $\mathbb{CP}^{N-1}$ Theory in the Presence of
Monopoles} \label{sec:CPN}

The insertion of one monopole into the partition function
of $\mathbb{CP}^{N-1}$ model in the disordered phase has been originally
considered in Ref.~\onlinecite{SachdevMurthy}. 
The $1/N$ expansion proceeds by replacing the quartic self-interactions in Eq.~(\ref{cpn}) by a fixed-length constraint
on the spinons; so we consider the action
\begin{equation}
\mathcal{S} = \int d^2 x d\tau \left[ |(\partial_\mu - i A_\mu ) z_\alpha |^2 + i \lambda \left( |z_\alpha|^2 - \frac{1}{g} 
\right) \right]
\end{equation}
where $\lambda$ is a fluctuating Lagrange multiplier field.
The procedure for generating the $1/N$ expansion is now simple.
One first integrates over
the $z$ fields obtaining an effective action for $A_{\mu}$ and $\lambda$. 
However, instead of expanding this effective action
around the trivial classical vacuum $A_{\mu} = 0$, one expands
around the monopole (instanton) solution, $A^i_{\mu}$, with \beq \label{Fmon}
F^i_{\mu} = 2 \pi q \frac{(x-x_0)_{\mu}}{4 \pi |x-x_0|^3}\eeq where
$F_{\mu} = \epsilon_{\mu \nu \lambda} \d_{\nu} A_{\lambda}$, $q$ is the monopole charge 
and $x_0$ is the monopole position. In practice, integrating out the $z$ fields in the background 
of spatially varying monopole fields is quite complicated (even more so due to the appearance of UV
 and IR divergences), so that only the leading term in the $1/N$ expansion has been computed in the past (that is fluctuations
 of $A_\mu$ about the monopole solution have not been taken into account). At this order, one finds,
 \beq \langle V^q(x) \rangle \sim \left(\frac{m}{\Lambda}\right)^{2 N \rho_q}\eeq
 where $V^q(x)$ is the monopole operator of charge $q$, $m$ is the mass gap of the theory, $\Lambda$ is the ultraviolet cut-off and $\rho_q$ is a collection of universal numbers (depending only on the charge of the monopole) which have been computed in Ref.~\onlinecite{SachdevMurthy}. Thus, the dimension of operator $V^q(x)$, $\mathrm{dim}[V^q] = 2 N \rho_q$.
 
If finding the expectation value of a monopole operator (and its scaling dimension) was very complicated, 
finding correlators of $V(x)$ with Wilson loops at $N=\infty$ turns out to be exceedingly simple. Indeed, we notice that
at leading order in $1/N$ it is sufficient to simply replace $A_{\mu}$ in the Wilson loop by its monopole value,
\beq \label{Vloop} \frac{\langle V^q(x) \exp\left(-i Q \int_{\cal C} A_{\mu} dx_{\mu}\right)\rangle}{\langle V^q \rangle} \to  \exp\left(-i Q \int_{\cal C} A^i_{\mu} dx_{\mu}\right)= \exp\left(-i Q \int_{\cal S} F^i_{\mu} dS_{\mu}\right)\eeq provided that we take the charge $Q$ to be $\mathcal{O}(1)$ in $N$ (otherwise, if $Q \sim \mathcal{O}(N)$ the Wilson line will change the background monopole field and the problem becomes intractable). Here ${\cal C}$ is some closed contour and ${\cal S}$ is any surface such that $\d{\cal S} = {\cal C}$. Thus, all we have to do is find the flux of our monopole through the Wilson loop that we are considering. Fluctuations of $A_{\mu}$ about the monopole field (\ref{Fmon}) will contribute at $\mathcal{O}(1/N)$ to the correlator (\ref{Vloop}). Likewise, if we denote the Wilson loop operator by $W({\cal C})$, then in the absence of the monopole field $\langle W({\cal C}) \rangle \sim 1+\mathcal{O}(1/N)$ (saturated by fluctuations of $A_{\mu}$ around the trivial vacuum), so
\beq \label{VW} \frac{\langle V^q(x) W({\cal C}) \rangle}{\langle W({\cal C}) \rangle} = \langle V^q \rangle \exp\left(-i Q \int_{\cal S} F^i_{\mu} dS_{\mu}\right)\eeq and at leading order in $1/N$ the external charge only changes the phase of the expectation value of monopole operator but not its magnitude.

In principle we are interested in finding the correlator of the monopole operator (that we place at a point $x = (r \cos\theta, r \sin\theta,0)$) and a straight, temporal Wilson line of charge $Q$ (which we place at the origin). However, to regularize possible $IR$ divergences let's also place a charge $-Q$ on the positive $x$ axis far away from the origin. As usual, we may connect the two oppositely directed Wilson lines in the far past and far future. Then,  according to (\ref{Vloop}) we have to compute the magnetic flux due to the monopole field (\ref{Fmon}) through the $y = 0, x > 0$ half-plane,
\bea \int \vec{F} \cdot d\vec{S} &=& -\frac{q}{2} \int_{-\infty}^{\infty} d\tau \int_0^{\infty} dx \frac{r \sin\theta}{((x-r\cos\theta)^2+r^2 \sin^2\theta + {\tau}^2)^\frac32}\\&=& -q \int_0^{\infty}dx \frac{r \sin\theta}{(x-r\cos\theta)^2+r^2\sin^2\theta} \\&=& -q (\pi - \theta)\eea
We see that the flux through the Wilson loop changes by $2 \pi q$ as the monopole crosses the surface of the loop. However, the expectation value,
\beq \label{vortex} \langle V^q(x) \rangle_{\mathrm{imp}}=\frac{\langle V^q(x) W({\cal C})\rangle}{\langle W({\cal C}) \rangle} = \langle V^q \rangle e^{i Q q (\pi - \theta)}\eeq
remains single valued, as by Dirac's condition $Q$ is an integer (in what follows, we shall also often discuss Wilson loops with non-integer charge $Q$, which in the presence of monopole operators are defined by specifying a surface ${\cal S}$, $W({\cal S}) = e^{-i Q \int_{\cal S} F_{\mu} dS_{\mu}}$. The correlation functions then explicitely depend on the choice of the surface, as can be seen from (\ref{vortex})). Thus, we see that the phase of the monopole operator winds by $-2 \pi Q$ as we move it in a full circle around the Wilson line, i.e. an external charge creates a vortex of the monopole field, consistent with the OPE in Eq.~ \ref{vimp5}. We expect that once we go beyond the leading order in $N$, this vortex will also get a nontrivial spatial profile,
\beq \label{vortex2}\langle V^q(x) \rangle_{\mathrm{imp}} = \langle V^q\rangle f(m |\vec{x}|) e^{-i Q q \theta} e^{i \chi} \eeq Here $f(r)$ is the vortex profile function and $e^{i \chi}$ is some overall phase (discussed below). We expect that far away from the external charge, the monopole field tends to its vacuum expectation value so that $f(\infty) = 1$. Moreover, by continuity we expect the monopole field to vanish at the origin, $f(0) = 0$. To the order to which we were working, $f(r) = 1$, which implies that the impurity exponent $\Delta^V_{\mathrm{imp}} \sim \mathcal{O}(1/N)$.

Notice that the result (\ref{vortex}) is sensitive to the angular position of the distant charge relative to the one  at the origin (we introduced the variable $\theta$ as the angle between the plane of the Wilson loop and the monopole operator). This is not unexpected: the monopole field is the order parameter for the flux symmetry, which is spontaneously broken in the disordered phase. As we rotate the distant charge, the overall phase $e^{i \chi}$ of the expectation value of the monopole operator changes - that is we explore different states in our vacuum manifold. 

If we were instead considering a correlation function of a string of monopole operators $\prod_i V^{q_i}(x_i)$ such that the overall combination is invariant under the flux symmetry (that is $\sum_i q_i = 0$) we expect the dependence on the angular position of the distant charge to drop out. We can check this in the limit $m |x_i - x_j| \gg 1$, $m |\vec{x}_i| \gg 1$, assuming clustering,
\beq \label{cluster} \langle \prod_i V^{q_i}(x_i)\rangle_{\mathrm{imp}} \to \prod_i \langle V^{q_i}(x_i) \rangle_{\mathrm{imp}} \to \prod_i \langle V^q_i \rangle e^{-i Q q_i \theta_i}\eeq
which is invariant under $\theta_i \to \theta_i + \chi$. Alternatively, in the same limit of far separated monopoles and at $N = \infty$, the classical magnetic field will just be a linear superposition of magnetic fields due to each monopole. Thus, the flux $\Phi$ through the Wilson loop will be given by, $\Phi = - \sum_i q_i (\pi - \theta_i)= \sum_i q_i \theta_i$ and using the equivalent of (\ref{VW}) for a string of monopole operators, we arive at the same expression (\ref{cluster}).

We expect the general form (\ref{vortex2}) to be preserved at any finite order in $1/N$. Nevertheless, in the flux-broken phase of the theory, there are also non-perturbative effects that should be taken into consideration. Indeed, the $U(1)_{\Phi}$ vortex nucleated by the external charge is global, and thus, will have a logarithmically divergent energy. Put into a more conventional language, the external charge creates a Coulomb potential, which is logarithmic in two dimensions, $V(r) \approx -\frac{e^2 Q}{2 \pi} \log(m r)$ for $m r \gg 1$. The effective coupling constant $e^2$ can be calculated in the $1/N$ expansion to be $e^2 \sim \frac{1}{N} m$. Thus, it will be energetically favourable for the external charge to bind a dynamical spinon (we concentrate on the case $Q = 1$ here for simplicity). This process can be analyzed by means of a non-relativistic Schrodinger equation\cite{KKLSS}. One finds a bound state of size $r_b \sim N^{\frac12} m^{-1}$. We expect that for $r \gg r_b$ the external charge will be screened by the dynamical spinon. On the other hand for $r \ll r_b$ this logarithmic confinement should generally have little effect on the physics. However, there is one notable exception: the expectation value of the monopole operator $V^q$ (\ref{vortex}) will be drastically altered on all distance scales by the screening. Indeed, if we assume that screening takes place, $\langle V^q(\vec{x}) \rangle$ has to tend to its vacuum expectation value for $|\vec{x}| \gg r_b$, and should experience no phase winding. We don't expect the winding number to change abruptly as we decrease $|\vec{x}|$, so we won't see a phase winding of $\langle V^q(\vec{x})\rangle$ on short distances $|\vec{x}| \ll r_b$ as well. 

A toy model for the disappearance of winding when screening effects are taken into account can be constructed as follows. We can use the charge $-Q$ that we previously put far away from the origin to represent the dynamical spinon that gets bound to the external charge. We first freeze the location of this spinon at some position $\vec{x}'$ away from the origin and compute the resulting expectation value of $V^q(\vec{x})$ using eq. (\ref{VW}). We then average the resulting $\langle V^q(\vec{x})\rangle$ over the spinon positions $x'$ with the probability distribution $|\psi(\vec{x}')|^2$, where $\psi(\vec{x})$ is the spinon wave-function. Since this wave-function will be azimuthally symmetric, one immediately learns that upon averaging over the angular position of the spinon, $\langle V^q(\vec{x})\rangle$ looses its finite winding number and will, in fact, carry a constant phase for all $\vec{x}$. This same averaging will also lead to an additional supression $\langle V^q(\vec{x})\rangle \sim |\vec{x}|$ as $\vec{x} \to 0$ (recall that at $N = \infty$ there was no supression of the vortex profile for $x \to 0$ before screening effects were taken into account). The origin of this supression is easy to see - for an external charge located infinitely far away, the averaging over the azimuthal position of the charge is identical to averaging of the phase $\chi$ in eq. (\ref{vortex2}) producing a zero result for $\langle V^q \rangle$. 

Do the above findings invalidate the OPE (\ref{vimp5})? The answer is no. The above discussion simply implies that $\langle V_{\mathrm{imp}} \rangle = 0$ and, thus, the expectation value $\langle V(\vec{x}) \rangle$ for $\vec{x} \to 0$ is controlled by higher order terms in the OPE (namely, by the impurity operator with angular momentum zero). However, higher correlation functions of $V$ operator, e.g. the $VBS$ susceptibility $\langle V(x) V^{\dagger}(x')\rangle$, are still controlled by the OPE (\ref{vimp5}). Such correlators are invariant under the $U(1)_{\Phi}$ symmetry, so as we argued above, their short distance properties are not sensitive to the location of the distant charge, and hence, to screening physics.

\section{Easy plane model at $N=1$}\label{sec:dualityN1}

This section, and the next, will examine a simplified version of the theory $\mathcal{S}+\mathcal{S}_{\rm imp}$
in which the non-abelian global SU($N$) symmetry is reduced to an abelian U(1)$^{N-1}$ symmetry. This enables us to use
the tools of abelian particle-vortex duality \cite{Peskin,Halpern} to obtain a theory expressed in terms of fields which are locally
related to the monopole/VBS operator $V$. The present section will consider the simplest case \cite{sj} with $N=1$. 
This model describes the onset of VBS order in a
$S=1/2$ quantum antiferromagnet in the presence of a staggered magnetic field \cite{senthil2}, and is useful in 
resolving a number of key technical questions in their simplest setting. 
For $N=1$, the theory $\mathcal{S}$ does not have any global
continuous symmetry, and becomes equivalent to scalar electrodynamics. With the results for the $N=1$ theory obtained
in the present section, we will be able to rapidly analyze the general $N$ case in the next section.

\subsection{Duality and Wilson loops}\label{sec:DualityW}
It is well known that in three space-time dimensions, near its critical point, non-compact $N=1$ scalar electrodynamics is dual to a theory of a complex (pseudo)scalar field with a global $U(1)$ symmetry \cite{Peskin,Halpern}. The Lagrangians of these two theories are as follows,
\bea \label{LQED}L_{QED} &=& \frac{1}{2 e^2} F^2_{\mu} + |(\d_{\mu} - i A_{\mu}) z|^2 + m^2 |z|^2 + \frac{g}{2} |z|^4 \\
\label{LXY}L_{XY} &=& |\d_{\mu} V|^2 + \tilde{m}^2 |V|^2 + \frac{\tilde g}{2} |V|^4\eea
Here $z$ and $V$ are complex one component fields. The duality is understood as being true for the range of parameters where $L_{QED}$ has a second order phase transition (which at weak coupling is believed to occur for $g/e^2$ sufficiently large). One way to understand the duality is by noting that the phase transition in scalar QED is driven by spontaneous breaking of flux symmetry $U(1)_\Phi$, which is precisely the global symmetry of $L_{XY}$. The order parameter for the flux symmery is the monopole operator $V(x)$ - that is the dynamical field of $L_{XY}$\cite{footnote1}. As we know, to each continous symmetry there corresponds a conserved current. In the case of flux symmetry of QED, this pseudo-vector current is just the magnetic field $F_{\mu}$, which is trivially conserved in the absence of monopoles, $\d_{\mu} F_{\mu} = 0$. Let's introduce an external field $H_{\mu}$ that would couple to this current,
\beq \delta L_{QED} = i H_{\mu} F_{\mu}\eeq  Suppose we are calculating some correlation function with insertion of a string of monopole operators of charge $q_i$ at points $x_i$. The gauge field $A_{\mu}$ in the path integral is then subject to the condition, $\d_{\mu} F_{\mu} = \sum_i 2\pi q_i \delta(x-x_i)$. Then under the transformation,
\beq \label{Htransf} H_\mu \to H_{\mu} + \d_{\mu} \alpha\eeq
\beq S_{QED} \to S_{QED} + i \int dx \,\d_{\mu} \alpha F_{\mu} = S_{QED} - i \int dx\, \alpha \,\d_{\mu} F_{\mu} = S_{QED} - 2 \pi i \sum_i q_i \alpha(x_i)\eeq
Hence, by introducing the field $H_{\mu}$ we can enlarge the global $U(1)_\Phi$ symmetry to a fictitious local symmetry, provided that the monopole operators transform as,
\beq \label{Vqtransf} V^{q}(x) \to e^{2 \pi i q \alpha(x)} V^q(x)\eeq 
%Thus, in any correlation function,
%\beq \langle \prod_i V^{q_i}(x_i) ...\rangle_{H_{\mu} + \d_{\mu} \alpha} = \langle \prod_i e^{i q_i \alpha(x_i)} %V^{q_i}(x_i) ...\rangle_{H_{\mu}}\eeq 
The dual Lagrangian $L_{XY}$ has to posses this local symmetry. Hence, to introduce the field $H_{\mu}$ into the dual Lagrangian we simply have to covariantize the derivative of the dynamical monopole field $V$,
\beq \label{covar} \d_{\mu} V \to D_{\mu} V = (\d_{\mu} - 2 \pi i H_{\mu}) V\eeq in eq. (\ref{LXY}). Other ``gauge invariant" operators can also be added to $L_{XY}$, e.g. $H^2_{\mu \nu}$; however, their contribution will, generally, either cancel out in correlation functions or be less singular near the critical point.

Thus, the dual Lagrangian in the presence of a background source field $H_{\mu}$ is given by,
\beq \label{dualH} L_{XY} = |(\d_{\mu}-2 \pi i H_{\mu}) V|^2 + \tilde{m}^2 |V|^2 + \frac{\tilde g}{2} |V|^4\eeq

The covariantization procedure (\ref{covar}) was explicitly written down in Ref.~\onlinecite{Laine}. Similar arguments for the case of a constant imaginary $H_{\mu}$, which physically represents an external magnetic field in the QED language and translates into a chemical potential for the flux symmetry in the XY language, have been given in Ref.~\onlinecite{Son}. In a companion paper\cite{MSQ} we shall also give an argument based on an exact duality transformation on the lattice, which will support (\ref{dualH}).

Having learned how to incorporate the source field $H_{\mu}$ into the dual Lagrangian, it is now trivial to dualize Wilson loops. Indeed, insertion of a Wilson loop $W({\cal C})$ into a correlation function is equivalent to adding into the Lagrangian the source term
\beq \delta L = i Q \int_{\cal C} dx_{\mu} A_{\mu} = i Q \int_{\cal S} dS_{\mu} F_{\mu} = i \int dx H_{\mu} F_{\mu} \eeq
where \beq \label{HWilson} H_{\mu}(x) = Q \int_{y \in {\cal S}} dS_\mu \,\delta(x-y)\eeq That is $H_{\mu}$ is a field that lives on the surface of the Wilson loop and is directed perpendicular to this surface.

Another benefit of introducing the source field $H_{\mu}$ is that by differentiating with respect to it we can compute correlation functions of the magnetic field $F_{\mu}$. For instance,
\beq \langle - i F_{\mu} (x) \rangle_H = \frac{\delta \log Z[H]}{\delta H_{\mu}(x)}=-2 \pi i \langle \left(V^{\dagger} D_{\mu} V - (D_{\mu} V)^{\dagger} V\right)(x)\rangle_H\eeq
That is the topological flux current $F_{\mu}$ of QED gets mapped into the Noether's current associated with the global $U(1)$ symmetry of the dual model. Differentiating once again,
\bea \langle F_{\mu}(x)  F_{\nu}(y)\rangle_{H,\,conn} &=& - \frac{\delta Z[H]}{\delta H_{\mu}(x) \delta H_{\nu}(y)}\\ &=& (2 \pi)^2 \left( \langle V^{\dagger} \overleftrightarrow{D}_{\mu} V
(x) V^{\dagger} \overleftrightarrow{D}_{\nu} V(y)\rangle_{H,\,conn} + 2 \delta_{\mu \nu} \delta(x-y) \langle V^{\dagger} V(x) \rangle_H\right)\nn\\ \label{FFcor}\eea The first term in (\ref{FFcor}) is the expected correlator of two $U(1)_{\Phi}$ currents, while the second term is a tadpole that ensures the overall transversality of the correlation function.

Having discussed the duality at length, we now return to our original problem: what is the influence of the exernal charge (Wilson line) on various physical observables. The observable of most interest to us is the monopole operator $V(x)$. However, this observable is physical only for integer-valued charge $Q$ of the Wilson line (Dirac's condition). Indeed, recall that in the dual language the field $H$ depends on a choice of surface ${\cal S}$ of the Wilson loop. If we pick a different surface ${\cal S}'$ then the field $H_{\mu}$ undergoes a gauge transformation $H_{\mu} \to H'_{\mu} = H_{\mu} + \d_{\mu} \alpha$ with $\alpha(x) = -Q\, 1_{x \in {\cal V}}$ where ${\cal V}$ is the volume bounded by the two surfaces ${\cal S}$ and ${\cal S}'$. Hence,
\beq \langle V(x) ... \rangle_{H'} = e^{2 \pi i \alpha(x)} \langle V(x) ... \rangle_{H}\eeq
where ellipses denote some other operators. Thus, the operator $V(x)$ is invariant under changing the surface of the Wilson loop if and only if $Q$ is an integer. However, if the charge $Q$ is a rational number, $Q = p/q$ where $p$ and $q$ are integers then the flux $2 \pi q$ monopole operator $V^q(x)\sim \left(V(x)\right)^q$ is physical. Moreover, a theory with arbitrary irrational $Q$ is still sensible provided that we confine our attention to correlation functions of operators which are invariant under the fictitious $U(1)_{\Phi}$ local symmetry, e.g. the magnetic field operator $-i F_{\mu} = - 2 \pi i V^{\dagger}\overleftrightarrow{D}_{\mu} V$. In fact, if we are dealing with such gauge invariant operators we don't necessarily have to use the precise form of $H$ given by (\ref{HWilson}); defining $\gamma_{\mu}$ to be a field living on the perimeter of the Wilson loop and directed along it, 
\beq \gamma_{\mu}(x) = Q \int_{y\in {\cal C}} dy_{\mu} \delta(x-y)\eeq we see that,
\beq \label{curlH} \epsilon_{\mu \nu \lambda} \d_{\nu} H_{\lambda} = \gamma_{\mu}\eeq
Then, by performing a suitable gauge transformation on $H_{\mu}$ and $V$ we can choose $H_{\mu}$ to be any field with curl given by $\gamma_{\mu}$. Thus, we see that the duality maps a Wilson loop of charge $Q$ in the QED language to an external magnetic flux tube of flux $2 \pi Q$ in the XY language. This correspondence has been noted in Ref.~\onlinecite{Samuel}, but the consequences of this correspondence for the critical properties of Wilson loops were not discussed.

%Similarly, if we are dealing with integer valued Wilson lines so that correlation functions of monopole operators are physical, we can still use any %$H$ satisfying the condition (\ref{curlH}) provided that we equip our monopole operators with ``dual Wilson lines," e.g. 
%\beq \label{dualWilson} \frac{\langle V(x) V^{\dagger}(y) W({\cal C})\rangle_{QED}}{\langle W({\cal C})\rangle_{QED}} = \langle V(x) e^{-2 \pi i \int_{\tilde{{\cal C}}} dx'_{\mu} H_{\mu}(x')} V^{\dagger}(y)\rangle_{H,\, XY}\eeq where $\tilde{\cal C}$ is any path stretching from $y$ to $x$.  Note that right hand side of (\ref{dualWilson}) is independent of the choice of the path $\tilde{\cal C}$ since integral Wilson lines get mapped to integral flux tubes, so $\oint dx_{\mu} H_{\mu}(x) = 2 \pi n, \, n \in {\mathbb Z}$.

%Finally, we note that correlation functions of monopole operators are mathematically well defined even for fractional charge $Q$, provided that we specify Wilson loops by their surface rather than by their perimeter.
%\subsection{Application of Duality}\label{sec:DualApp}
Now we can address the problem that we originally posed in a dual language. Let's place an charge external charge $Q$ at the spatial origin. For now we don't insist that this charge be an integer. The dual source field $H_{\mu}$ must, therefore, satisfy
\beq \label{rotH} \nabla \times \vec{H} =  Q \delta^2(\vec{x}) \hat{\tau}\eeq  Thus, we basically have to solve an Aharonov-Bohm problem with flux $2 \pi Q$. One choice for the source field $H_{\mu}$ is
\beq \label{stringH} H_{\mu}(x) = Q \delta_{\mu, 2} \theta(x) \delta(y)\eeq
This is the so-called string gauge, which corresponds to (\ref{HWilson}), with the surface of the Wilson loop being the plane $y = 0$, $x > 0$. As is well known, the string gauge is equivalent to $H_{\mu} = 0$ and the boundary condition,
\beq \label{bc} V(\theta = 2 \pi) = e^{-2 \pi i Q} V(\theta = 0)\eeq
where $\theta$ is the azimuthal angle. Thus, we have to solve the theory (\ref{LXY}) with the twisted boundary condition (\ref{bc}). We observe that the physics is, therefore, a periodic function of $Q$. For integer $Q$ the boundary condition (\ref{bc}) is trivial - there is no twist. So our argument indicates that integral external charges do not affect correlation functions on distances of order of the correlation length of the theory: screening of integral charges takes place on distance scales of oder of microscopic $UV$ cutoff. This surprising fact is discussed in more detail in a companion paper\cite{MSQ}. 

The behaviour at non-integer $Q$ is less unexpected. One physical question that we may ask is what is the magnetic (electric) field induced by the charge $Q$ (we define the electric field $E_i = F_{i3} = - \epsilon_{ij} F_j$ where latin letters $i,j,k$ run over spatial indices). Although this is a departure from our original goal, we will see that a lot of the results that we will obtain along the way will be useful when we return to discuss correlators of monopole field for the planar theory with $N$ fields. Another question that we will adress for non-integer, rational, values of $Q = p/q$ is the behaviour of higher flux monopole operators $V(x)^q$.

\subsection{Perturbative expansion of the dual theory for $Q \to 0$}\label{sec:Qzero}
The magnetic field $-i F_{\mu}$ is a conserved current and receives no renormalizations and, thus, has conformal dimension $2$. Therefore, at the critical point we expect,
\beq \label{EfieldQ} \langle -i \vec{E} \rangle = C(Q) \frac{1}{r^2} \hat{r}\eeq
The electric field is imaginary as we are working in Euclidean space. The coeficient $C(Q)$ is a universal number that is a periodic function of charge $Q$. We shall be interested in determining this function.

For $Q \to 0$ we can perform a perturbative expansion in $H_{\mu} \sim \mathcal{O}(Q)$. 
\beq \label{FieldPert} \langle - i F_{\mu}(x) \rangle_H =  \frac{\delta \log Z[H]}{\delta H_{\mu}(x)} \approx \int dy \frac{\delta^2 \log Z}{\delta H_{\mu}(x) \delta H_{\nu} (y)} H_{\nu} (y) = - \int dy \langle F_{\mu}(x) F_{\nu} (y) \rangle H_{\nu}(y)\eeq
As we have learned, the correlation function of magnetic field $F_{\mu}$ dualizes to,
\beq K_{\mu \nu}(x-y) = \langle F_{\mu} (x) F_{\nu} (y)\rangle = (2 \pi)^2 \left( \langle V^{\dagger} \overleftrightarrow{\d_{\mu}} V
(x) V^{\dagger} \overleftrightarrow{\d_{\nu}} V(y)\rangle + 2 \delta_{\mu \nu} \delta(x-y) \langle V^{\dagger} V \rangle\right)\eeq By transversality,
\beq K_{\mu \nu}(p) = K(p) (\delta_{\mu \nu} - \frac{p_{\mu} p_{\nu}}{p^2})\eeq
By RG $K(p)$ should have the form,
\beq K(p) = M g(p/M) \eeq where $M$ is some physical scale in the theory (e.g. in the $U(1)_{\Phi}$ disordered phase, the mass of the monopole field $V$). At the critical point,
\beq \label{Kcrit} K(p) = A |p| \eeq
where $A$ is some universal number. On the XY  side of the theory, this universal number has been computed before using both $\epsilon$ expansion\cite{Fazio} and large $M$ expansion\cite{FisherGirvin}. The large $M$ expansion is obtained by replacing the complex scalar $V$ in the action for the XY theory (\ref{LXY}) by an $M$ component complex field. In the large $M$ expansion the coefficient $A$ is found to be at next to leading order in $M$,
\beq \label{AN} A = (2 \pi)^2 \frac{M}{16}\left(1-\frac{1}{M} \frac{32}{9 \pi^2}\right) \stackrel{M=1}{\approx} 1.6\eeq
while in the $\epsilon$ expansion one obtains $A \approx 2.0$ at $\mathcal{O}(\epsilon^2$). Monte-Carlo simulations on the XY model\cite{FisherGirvin} indicate  $A \approx 1.8$.\cite{footnote2} The coefficient $A$ can also be computed by performing a large $N$ expansion in the original QED, whereby the field $z$ is promoted to have $N$ components. At leading order one obtains $A = 16/N \stackrel{N=1}{=} 16$ (as usual, direct large $N$ expansion in QED produces results, which are numerically notoriously inaccurate for $N \sim 1$).

For completeness, we also discuss the behaviour of $K(p)$ at small momenta on both sides of the critical point. In the phase where the $U(1)_\Phi$ symmetry is spontaneously broken the spectrum of the theory should contain a goldstone, which can be created out of the vacuum by the $U(1)_\Phi$ current,
\beq \label{matrixel} \lim_{p\to 0} \langle p|F_{\mu}(x)|0\rangle = 2 \pi \lim_{p \to 0} \langle p|-i V^{\dagger} \overleftrightarrow{\d_{\mu}} V(x)|0\rangle  = 2 \pi i f p_{\mu} e^{i p x}\eeq where in three dimensions $f^2$ defines a physical energy scale. Note that equation (\ref{matrixel}) is written in Minkowski space. We see that the goldstone is nothing but the photon of the original QED. Then $K_{\mu \nu}(p)$ should have a pole at $p^2 = 0$ and using spectral decomposition, 
\beq \lim_{p\to 0} K(p) = (2 \pi f)^2 \eeq
On the other hand, in the the phase where the $U(1)_{\Phi}$ symmetry is unbroken (that is in the ``superconducting" phase of QED) the $V$ field is massive and all the excitatations have a gap. Therefore, $K_{\mu \nu}(p)$ cannot have a pole at $p^2 = 0$ and
\beq \lim_{p \to 0} K(p) \sim \frac{p^2}{M}\eeq
Having discussed the expected form of $K_{\mu \nu}$ in different phases we can go back to eq. (\ref{FieldPert}) for electric field induced by the charge $Q$. Introducing the kernel ${\cal D}(p) = K(p)/p^2$, and using eq. (\ref{rotH}),
\beq \langle - i F_{\mu}(x)\rangle = - \int dy K_{\mu \nu}(x-y) H_{\nu}(y) = - Q \int d \tau' \epsilon_{\mu \nu 3} \d^{x}_{\nu} {\cal D}(\vec{x},\tau')\eeq Hence,
\beq \langle - i \vec{E}(\vec{x}) \rangle = Q h(|\vec{x}|) \hat{r}\eeq
where \beq h(|\vec{x}|) = -\frac{\d}{\d |\vec{x}|} \int d\tau'{\cal D}(\vec{x},\tau')\eeq
Substituting the expression (\ref{Kcrit}) for $K(p)$ at the critical point we obtain,
\beq \langle - i \vec{E}(\vec{x}) \rangle = Q \, \frac{ A}{2 \pi |\vec{x}|^2} \hat{r}\eeq
Hence we identify,
\beq \label{CAconn} C(Q) \approx Q A/(2 \pi),  \quad Q \to 0\eeq 
Similarly, in the $U(1)_\Phi$ ordered phase,
\beq \langle - i \vec{E}(\vec{x}) \rangle = Q \frac{ 2 \pi f^2}{|\vec{x}|} \hat{r}\eeq
So in this phase, as expected, the external electric charge produces the usual Coulomb-like electric field, $\vec{E} = \frac{e_{eff}^2 Q}{2 \pi r}$, as appropriate to two spatial dimensions with the identification $e_{eff} = 2 \pi f$.

\subsection{Peculiarities of the free theory}\label{sec:FreeTheory}
So far we have only discussed the leading term in $C(Q)$ for $Q \to 0$. In principle, we could continue the expansion in $Q$ to higher orders: then the problem reduces to finding correlators of current operators $-i F_{\mu} = i V^{\dagger} \overleftrightarrow{\d_{\mu} } V$. These correlators can be found by performing either $\epsilon$ or $1/M$ expansion of the XY model. In either case, going beyond the leading order in $Q$ is not simple. So, instead, we choose to return to the formulation of the problem involving the twisted boundary condition (\ref{bc}). In the next section we will use this formulation to compute $C(Q)$ for all $Q$ (albeit numerically) at $M= \infty$. However, before we do so, we will solve a slightly simpler problem: namely we find the form of $C(Q)$ at the gaussian fixed point $\tilde g = 0$, $\tilde{m}^2 = 0$ of the Lagrangian (\ref{LXY}). The reason for studying the free theory is that the calculations in it are, technically, very similar to those in the strongly coupled $M = \infty$ theory addressed in the next section (even though the physical results are quite different).

In the free theory, $C(Q)$ can be determined exactly, and, surprisingly, turns out to be a non-analytic function of $Q$ at $Q=0$. We have not been able to see any hints of this non-analyticity from the perturbative expansion of the free theory in $Q$ (perhaps because we could go perturbatively only to linear order in $Q$, whereas the non-analyticity of $C(Q)$ starts only at order $|Q|^2$). On the other hand, once we go in the next section to the strongly interacting fixed point (obtained in the $M = \infty$ limit), the theory cures itself of all $IR$ divergences and $C(Q)$ becomes analytic in $Q$.

So, let's compute,
\beq \label{diffprop}\langle -i F_{\mu}(x) \rangle = \langle -2 \pi i V^{\dagger}\overleftrightarrow{\d_{\mu}}V(x)\rangle=-2 \pi i \lim_{x \to y}(\d^x_{\mu}-\d^y_{\mu})\langle V(x) V^{\dagger}(y)\rangle\eeq in the free theory, $L = |\d_{\mu} V|^2$ subject to boundary condition (\ref{bc}). As eq. (\ref{diffprop}) shows, to find the $U(1)_{\Phi}$ current it is sufficient to determine the propagator, $D(x-y) = \langle V(x) V^\dagger(y)\rangle$.  The propagator will also determine the correlation function of operators $\left(V(x)\right)^q$ for rational $Q = p/q$,
\beq \label{VqVqfree} \langle \left(V(x)\right)^q \left(V^{\dagger}(y)\right)^q\rangle = q! D(x-y)^q\eeq We note that our problem is invariant under translations along the temporal direction, so,
\beq \label{D2plus1}D(\vec{x},\vec{x}',\tau-\tau') = \int \frac{d \omega}{2 \pi} D_2(\vec{x},\vec{x}',{\omega}^2) e^{i \omega (\tau-\tau')}\eeq
where $D_2(\vec{x},\vec{x}',{\omega}^2)$ denotes the two-dimensional propagator with mass $m^2 = {\omega}^2$ and twisted b.c. (\ref{bc}). We use spectral decomposition to find  $D_2$,
\beq \label{D2Spec}D_2(\vec{x},\vec{x}',m^2) = \sum_l \frac{e^{i l \theta}}{2 \pi}\int_0^{\infty} dE \frac{1}{m^2 + E} \phi_{l,E}(\vec{r}) \phi^*_{l,E}(\vec{r}') \eeq where we sum over states with fixed azimuthal angular momentum $l = n-Q$, $n \in {\mathbb Z}$. Note that the angular momenta are not integral due to the twisted b.c. (\ref{bc}). The radial eigenfunctions $\phi_{l,E}(r)$ satisfy,
\beq \label{ODEfree} \left(-\frac{1}{r}\frac{\d}{\d r}(r \frac{\d}{\d r}) + \frac{l^2}{r^2}\right) \phi_{l,E}(r) = E \phi_{l,E}(r)\eeq
and are normalized as,
\beq \int_0^{\infty} dr \,r \phi^*_{l,E}(r) \phi_{l,E'}(r) = \delta(E-E')\eeq
The solution to ODE (\ref{ODEfree}) is,
\beq \label{BesselFree}\phi_{l,E}(r) = \frac{1}{\sqrt{2}} J_{|l|}(\sqrt{E} r)\eeq
where $J_n(u)$ is the n-th order Bessel function. Hence,
\beq D(r,r',\theta-\theta',\tau-\tau') = \sum_{l} e^{i l (\theta-\theta')} \int \frac{d \omega}{2\pi} e^{i \omega (\tau-\tau')} \int_0^{\infty} \frac{du}{2 \pi} \frac{u}{u^2+\omega^2} J_{|l|}(u r) J_{|l|}(u r')\eeq where we made the substitution $u = \sqrt{E}$. Integrating over $\omega$,
\beq \label{DSpecFree}D(r,r',\theta,\tau) = \frac{1}{4 \pi r'} \sum_{l} e^{i l \theta} \int_0^{\infty} dv J_{|l|}(\frac{r}{r'}v)J_{|l|}(v) \exp(- \frac{|\tau|}{r'} v)\eeq
Now we can ask, what is the behaviour of the propagator $D(r,r',\theta,\tau)$ for $r \to 0$, i.e. for $r \ll r'$.  Recalling, $J_{|l|}(r) \approx \frac{1}{2^{|l|} \Gamma(|l|+1)} r^{|l|}$,
\beq \int_0^{\infty} dv J_{|l|}(\frac{r}{r'}v)J_{|l|}(v) \exp(- \frac{|\tau|}{r'} v)\approx  \left(\frac{r}{r'}\right)^{|l|} B_l(\frac{|\tau|}{r'})\eeq
with \beq B_l(u) = \frac{1}{2^{|l|} \Gamma(|l|+1)}\int dv v^{|l|} J_{|l|}(v) \exp(-u v)= \frac{\Gamma(|l|+\frac12)}{\Gamma(|l|+1)}\left(1+\frac{{\tau}^2}{r'^2}\right)^{-|l|-\frac12}\eeq
Thus, for $r \to 0$ the contribution of states with angular momentum $l$ to the propagator scales as $r^{|l|}$. So, the largest contribution comes from smallest $|l| = |n-Q|$. For $-\frac12 < Q < \frac12$ smallest $|l|$ is given by setting $n = 0$, $l = -Q$. Hence, for $|Q| < 1/2$, and $r/r' \ll 1$, 
\beq \label{assymDfree} D(r,r',\theta,\tau) \approx \frac{1}{4 \pi r'} \left(\frac{r}{r'}\right)^{|Q|} e^{-i Q \theta} B_Q(\frac{\tau}{r'})\eeq 
For values of $|Q| > 1/2$ we simply periodize the eq. (\ref{assymDfree}), since all physics in XY model is periodic in $Q$ with period $1$ (see discussion in previous section). From here on, we therefore confine our attention to $|Q| < 1/2$. 

Thus, if we were to perform the OPE in Eq.~(\ref{vimp5}) in the XY model
\beq \label{OPEfree}V(\vec{x}, \tau) \sim |\vec{x}|^{\Delta^V_{\mathrm{imp}}} e^{-i Q \theta} \, V_{\mathrm{imp}}(\tau)\quad \mathrm{for}\,\, |\vec{x}| \to 0\eeq
we would obtain for $|Q| < \frac12$ in the free XY model,
\beq \label{DeltaVFree}\Delta^V_{\mathrm{imp}} = |Q|. \eeq We immediately see that the free theory is non-analytic in $Q$ at $Q = 0$. By periodizing in $Q$, we also see that $\Delta^V_{\mathrm{imp}}$ is non-analytic at $Q = \pm 1/2$. However, this later non-analyticity appears only after we take $r \to 0$ limit of the propagator, while we expect the non-analyticity at $Q = 0$ to persist in the propagator for arbitrary $r,r'$. 

In fact, $Q = \frac{1}{2}$ is a very special point. At this point the $n = 0$, $l = -Q$ and $n = 1$, $l = 1-Q$, i.e. $l = \pm 1/2$ terms in the sum (\ref{DSpecFree}) become equally important for $r/r' \to 0$. Thus, for $Q \to 1/2$ it makes sense to keep both terms in the assymptotic expansion of the propagator,
\beq D(r,r',\theta,\tau) \approx \frac{1}{4 \pi r'}\left( \left(\frac{r}{r'}\right)^{Q} e^{-i Q \theta} B_Q(\frac{\tau}{r'}) + \left(\frac{r}{r'}\right)^{1-Q} e^{-i (Q-1) \theta} B_{Q-1}(\frac{\tau}{r'})\right)\eeq
and we may hypothesize the impurity OPE, for $Q \to 1/2$,
\beq \label{OPEdeg}V(\vec{x}, \tau) \sim  c_Q |\vec{x}|^{\Delta^V_{Q}} e^{-i Q \theta} \, V_{Q}(\tau) + c_{Q-1} |\vec{x}|^{\Delta^V_{Q-1}} e^{-i (Q- 1) \theta} \, V_{Q-1}(\tau), \quad \mathrm{for}\,\, |\vec{x}| \to 0\eeq
where $V_Q$ and $V_{Q-1}$ are two impurity operators, with impurity anomalous dimensions $\Delta^V_Q$ and $\Delta^V_{Q-1}$. In the free theory, $\Delta^V_{Q} = Q$ and $\Delta^V_{Q-1} = 1-Q$. Hence, for $Q < 1/2$, $\Delta^V_Q < \Delta^{V}_{Q-1}$ and the operator $V_Q$ is the most relevant as $|\vec{x}| \to 0$, while the operator $V_{Q-1}$ provides a subleading correction. For $Q > 1/2$ the roles of these two operators are reversed. Finally, for $Q = 1/2$ the two operators have degenerate anomalous dimensions, $\Delta^V_{1/2} = \Delta^V_{-1/2}$ and,
\beq \label{OPEdeg2}V(\vec{x}, \tau) \sim  c_{1/2} |\vec{x}|^{\Delta^V_{1/2}} e^{-i \theta/2} \, V_{1/2}(\tau) + c_{-1/2} |\vec{x}|^{\Delta^V_{-1/2}} e^{i \theta/2} \, V_{-1/2}(\tau), \quad \mathrm{for}\,\, |\vec{x}| \to 0\eeq

Physically, the $Q = 1/2$ point is special because the CP symmetry is effectively restored at it\cite{footnote3}. Indeed, under CP, $Q \to -Q$. However, as already discussed, the universal physics is periodic in $Q$, so the points $Q = \pm 1/2$ are identified. Thus, the two impurity operators, $V_{\pm 1/2}$ are just CP conjugates of each other and must have the same impurity anomalous dimensions. Hence, although our original analysis was performed for the case of the free theory, we expect the conclusions to remain valid in the strongly interacting theory.

We remind the reader that even though the operator $V(x)$ is mathematically well defined by specifying the surface ${\cal S}$ of the Wilson loop for arbitrary $Q$, it is not physical for non-integral $Q$. Indeed, a physical operator cannot obey twisted boundary conditions. However, for rational $Q = p/q$, the flux $2 \pi q$ monopole operator $V^q(x) \sim \left(V(x)\right)^q$ is well-defined on both sides of the duality. Using (\ref{VqVqfree}) and (\ref{assymDfree}), we obtain the OPE,
\beq \label{OPEVq} V^q(\vec{x}, \tau) \sim |\vec{x}|^{\Delta^{V}_{\mathrm{imp}}(q)} e^{-i q Q \theta} \, V^q_{\mathrm{imp}}(\tau)\quad \mathrm{for}\,\, |\vec{x}| \to 0\eeq
with
\beq \Delta^V_{\mathrm{imp}}(q) = q |Q|\eeq
in the free XY theory for $|Q| < 1/2$. Since $q Q = p$ is an integer, the OPE (\ref{OPEVq}) is invariant under $\theta \to \theta + 2\pi$, making the operator $V^q(x)$ single-valued, as required. 

Having discussed the impurity OPEs, let us return to the calculation of electric field. Since we know that the electric field will be radial, we only need the $\hat{\theta}$ component of the magnetic field,
\beq \label{FieldProp} \langle - i F_{\theta} \rangle = - 2 \pi i \frac{1}{r} \lim_{\theta \to \theta'} \left(\d_\theta - \d_{\theta'}\right) D(r=r', \theta - \theta', \tau = \tau') = - 4 \pi i \frac{1}{r} \lim_{\theta \to 0} \d_\theta D(r=r',\theta, \tau=\tau')\eeq
For this purpose, we don't need the propagator with $r/r' \ll 1$, but rather with $r \to r'$, $\tau \to \tau'$. We denote, $D(r,\theta) = D(r=r',\theta,\tau =\tau')$. Unfortunately, if we plug $r = r'$, $\tau = \tau'$ into the expression for propagator (\ref{DSpecFree}), the integral over $v$ diverges. We expect that if we instead first keep $r-r'$, $\tau-\tau'$ finite, perform the integration over $v$, sum over angular momenta $l$ and only then take $r = r'$, $\tau = \tau'$, the divergence disappears. There are also other ways to regularize the propagator: e.g. make the integral over $\omega$ in (\ref{D2plus1}) run over $D-2$ dimensions. This would correspond to the XY model in $D$ dimensions coupled to an external flux-tube (the flux-tube is a defect in $2$ dimensions, so its world-volume is $D-2$ dimensional). One then takes the limit $D \to 3$ at the end of the calculation. We have successfully used this method to compute the electric field (see Appendix \ref{AppFieldFree}). The result for the coefficient $C(Q)$ of eq. (\ref{EfieldQ}) is,
\beq \label{CQfree} C(Q) = \frac{1}{8} (1-2|Q|)^2 \tan(\pi Q), \quad |Q| < 1\eeq
Thus, we see that the function $C(Q)$ is non-analytic at $Q =0$. This analyticity occurs at non-leading order in $Q$,
\beq \label{CQfreeassym} C(Q) \approx \frac{\pi}{8} \, Q (1-4|Q|), \quad Q \to 0\eeq
The leading order term, $C(Q) \approx \frac{\pi}{8} \, Q$ is the one which would have been predicted by expanding the free theory perturbatively in $Q$.

One can also derive the result (\ref{CQfree}) in a different way, which can be more easily generalized from the free theory to the $1/M$ expansion in a strongly interacting theory. This calculation is based on the integral representation of the propagator of the twisted theory derived in Ref.~\onlinecite{Linet}. We repeat the calculations of Ref.~\onlinecite{Linet} in Appendix \ref{AppLinetProp} as in the next section we will need to generalize them for application in $1/M$ expansion. The result is,
\beq \label{DFreeInt} D(r,\theta) = \frac{1}{4 \pi r} \int_0^{\infty} d\nu \tanh(\pi \nu) U_{\nu}(\theta)\eeq
with, 
\beq \label{Ufull}U_{\nu}(\theta) = \frac{e^{- 2 \pi i Q \mathrm{sgn}(\theta)} \sinh(\nu |\theta|) + \sinh(\nu (2\pi -|\theta|))}{\cosh(2 \pi \nu)-\cos(2 \pi Q)}\eeq
from which one recovers eq. (\ref{CQfree}) by using eq. (\ref{FieldProp}), see Appendix \ref{AppLinetProp}.

\subsection{$1/M$ expansion of the dual theory}\label{sec:dualLargeN}
We now progress from the free XY model to the $1/M$ expansion of the strongly interacting theory. We take the Lagrangian to be,
\beq \label{LXYN} L = |\d_{\mu} V|^2 + i \lambda (|V|^2 - \frac{1}{g})\eeq
Here $V$ is an $M$ component complex scalar and $\lambda$ is a Lagrange multiplier, which enforces the local constraint,
\beq \label{constr} |V|^2 = \frac{1}{g}\eeq This hard constraint replaces the self-interaction of the $V$ field. In the presence of an external charge in the direct theory, we take $V$ to satisfy the twisted boundary conditions (\ref{bc}). In principle, we would like to solve the theory (\ref{LXYN}) in the limit $M \to 1$. However, practically we will only be able to perform computations at $M = \infty$.  

We will be interested in the properties of the theory (\ref{bc}) at its critical point $g = g_c$. As is well known from standard $1/M$ expansion techniques, at $M = \infty$ the critical coupling is given by,
\beq \label{gc} \frac{1}{M g_c} = \frac{1}{M} \langle V^{\dagger} V \rangle = D(x=x')\eeq
where $D$ is the usual massless 3D propagator,
\beq D(x, x') = \frac{1}{4 \pi |x-x'|}\eeq
Of course, the propagator with $x = x'$ in (\ref{gc}) is $UV$ singular and has to be regularized. Since we will perform calculations of propagator in position space, it is convenient for us to use point-splitting regularization.

In the absence of the twisted boundary condition (\ref{bc}) and at the critical point, we perform the expansion around $\langle i \lambda \rangle = 0$ (so that the effective mass for the $V$ particles vanishes). However, once $Q$ is finite, $\lambda = 0$ is no longer sufficient to make the constraint (\ref{constr}) satisfied. Instead, the Lagrange muliplier aquires a spatial dependence 
\beq \label{lambdaaQ}\langle i \lambda (\vec{x}, \tau)\rangle = \frac{a(Q)}{|\vec{x}|^2}\eeq
Here $a$ is a universal function of the charge $Q$. The dependence on $\vec{x}$ is determined from the canonical dimension of $\lambda$ ($\lambda$ aquires a non-trivial anomalous dimension only at order $1/M$). Thus, at finite $Q$, the propagator of $V$ field satisfies,
\beq \label{PropEqa} (-\d^2 + \frac{a(Q)}{|\vec{x}|^2}) D(x,x',Q) = \delta(x-x')\eeq
and $a(Q)$ should be determined self-consistently from the equation,
\beq \label{gapQ} \frac{1}{M g_c} = \frac{1}{M} \langle V^{\dagger} V\rangle_Q = D(x=x',Q)\eeq
Combining eqs. (\ref{gc}), (\ref{gapQ}),
\beq \label{gapQdif}\lim_{x\to x'} \left(D(x,x',Q) - D(x,x',Q=0)\right) = 0\eeq

Thus, the problem is reduced to finding the propagator $D(x,x',Q)$.  Just as in the free case, we use spectal decomposition (\ref{D2Spec}), and the radial functions $\phi_{l,E}(r)$ now satisfy,
\beq \label{eigensN} \left(-\frac{1}{r} \frac{\d}{\d r}(r \frac{\d}{\d r}) +\frac{l^2 + a}{r^2}\right) \phi_{l,E}(r) = E \phi_{l,E}(r)\eeq
where again due to the twisted boundary conditions $l = n - Q$, $n \in {\mathbb Z}$.
The solution to (\ref{eigensN}) is,
\beq \phi_{l,E}(r) = \frac{1}{\sqrt{2}} J_{\sqrt{l^2+a}}(\sqrt{E}r)\eeq
Comparing the result above to free theory (\ref{BesselFree}), we see that the only difference is in the replacement of the indices of Bessel functions $|l| \to \sqrt{l^2+a}$. Going from $2D$ to $3D$ propagator as in the free case (\ref{DSpecFree}),
\beq \label{DSpecN}D(r,r',\theta,\tau) = \frac{1}{4 \pi r'} \sum_{l} e^{i l \theta} \int_0^{\infty} dv J_{\sqrt{l^2+a}}(\frac{r}{r'}v)J_{\sqrt{l^2+a}}(v) \exp(- \frac{|\tau|}{r'} v)\eeq
Finally, expanding the propagator (\ref{DSpecN}) for $r \ll r'$, we obtain the equivalent of (\ref{assymDfree}),
\beq \label{assymDN} D(r,r',\theta,\tau) \approx \frac{1}{4 \pi r'} \left(\frac{r}{r'}\right)^{\sqrt{Q^2+a(Q)}} e^{-i Q \theta} B_{\sqrt{Q^2+a(Q)}}(\frac{\tau}{r'}), \quad |Q| < 1/2\eeq 
Thus, we recover the OPE (\ref{OPEfree}), but the impurity exponent now becomes some nontrivial function of $Q$,
\beq \label{DeltaVN1M}\Delta^V_{\mathrm{imp}} = \sqrt{Q^2+a(Q)}, \quad |Q| < 1/2\eeq
We note that, similar to the free case, as $Q$ passes $1/2$, the most relevant angular momentum $l$ in the sum (\ref{DSpecN}) changes from $l = -1/2$ to $l = 1/2$, and at $Q = 1/2$ we have the OPE (\ref{OPEdeg2}) with two degenerate impurity operators.

To find the nontrivial impurity exponent we need to solve eq. (\ref{gapQdif}) for $a(Q)$. We are, therefore, after the propagator $D(x,x',Q)$ with $x \to x'$. We could, in principle proceed as in the free case. Namely, make our flux-tube uniform along $D-2$ spatial dimensions (introducing a convergence factor $v^{D-3}$ into (\ref{DSpecN})), perform the integrals in (\ref{DSpecN}) with $r=r'$, $\tau = 0$, perform the sum over the angular momenta $l$, take $\theta \to 0$ and $D \to 3$. However, unlike in the free case, the sums over angular momenta cannot be now performed analytically in terms of hypergeometric functions (with nice analytic continuation for $\theta \to 0$). The sum over $l$ can still be performed numerically, however, the convergence is rather slow. Nevertheless, we have been able to determine $a(Q)$ numerically using this method.
However, this method is less suitable for finding the electric field coefficient $C(Q)$, which requires us to differentiate the propagator at $\theta = 0$, making the convergence properties of the series even worse.

Instead, we shall use a different method, generalizing the integral form of the propagator (\ref{DFreeInt}) derived in Ref.~\onlinecite{Linet} to the present problem. As shown in Appendix \ref{AppLinetProp}, the twisted propagator at $M = \infty$ is given by,
\beq \label{DintN} D(r,\theta) = \frac{1}{4 \pi r} \int_0^{\infty} d\nu \tanh(\pi \nu) \frac{\nu}{\sqrt{\nu^2+a}} U_{\sqrt{\nu^2 + a}}(\theta)\eeq
with $U_{\nu}(\theta)$ still given by eq. (\ref{Ufull}).

Now, $a(Q)$ can be determined from (\ref{gapQdif}),
\bea \label{gapQa} 0 &=& \lim_{\theta \to 0}(D(r,\theta,Q)-D(r,\theta,Q=0))\\ &=&\frac{1}{4 \pi r} \int_0^{\infty} d\nu \tanh(\pi \nu) \left(\frac{\nu}{\sqrt{\nu^2+a}}\frac{\sinh(2 \pi \sqrt{\nu^2+a})}{\cosh(2 \pi \sqrt{\nu^2+a})-\cos(2\pi Q)}-\frac{\sinh(2 \pi \nu)}{\cosh(2 \pi \nu) -1}\right)\nn\\\eea
Eq. (\ref{gapQa}) can be solved numerically for $a(Q)$. However, before we do this, let's verify our claim that $\langle i \lambda\rangle = 0$ (i.e. $a = 0$) is not sufficient to satisfy (\ref{gapQ}) for finite $Q$. Indeed, from (\ref{gapQa}) we obtain
\bea &&\lim_{x\to x'}\left(D(x,x',Q,a=0) -D(x,x',Q=0)\right) = \frac{1}{M} \left( \langle V^{\dagger} V(x)\rangle_{Q} - \langle V^{\dagger} V(x)\rangle_{Q=0}\right)\nn\\&=&\frac{1}{4 \pi r} \int_0^{\infty} d\nu \frac{\cos(2 \pi Q)-1}{\cosh(2 \pi \nu) - \cos(2 \pi Q)}= -\frac{1}{8 \pi r} (1-2 |Q|) \tan(\pi |Q|),\quad |Q| < 1\nn\\\label{VVdeficit2}\eea
where expectation values in the first line of (\ref{VVdeficit2}) are computed in the free theory. The precise value of expression (\ref{VVdeficit2}) is not very important for our purposes (although it is curious to note that like many quantities in the free theory it is non-analytic in $Q$ at $Q=0$). What is important for us is that expression (\ref{VVdeficit2}) is negative. This means that the twisted boundary condition effectively creates a repulsive barrier, leading to a decrease in $V^{\dagger} V$ compared to untwisted theory. To compensate for this decrease in the strongly interacting theory, we need $\langle i \lambda(x)\rangle$ to provide an attractive potential for $V$ particles. Hence, we conclude that $a(Q) < 0$ for $Q$ finite. One may be concerned that the square roots in expressions (\ref{DintN}), (\ref{gapQa}) are ambigious for $a < 0$ and $\nu^2 < |a|$. However, it turns out that these expressions do not depend on our choice of the sign for the square root as long as it is consistent. 
\begin{figure}[t]
\begin{center}
\includegraphics[angle=-90,width = 0.8\textwidth]{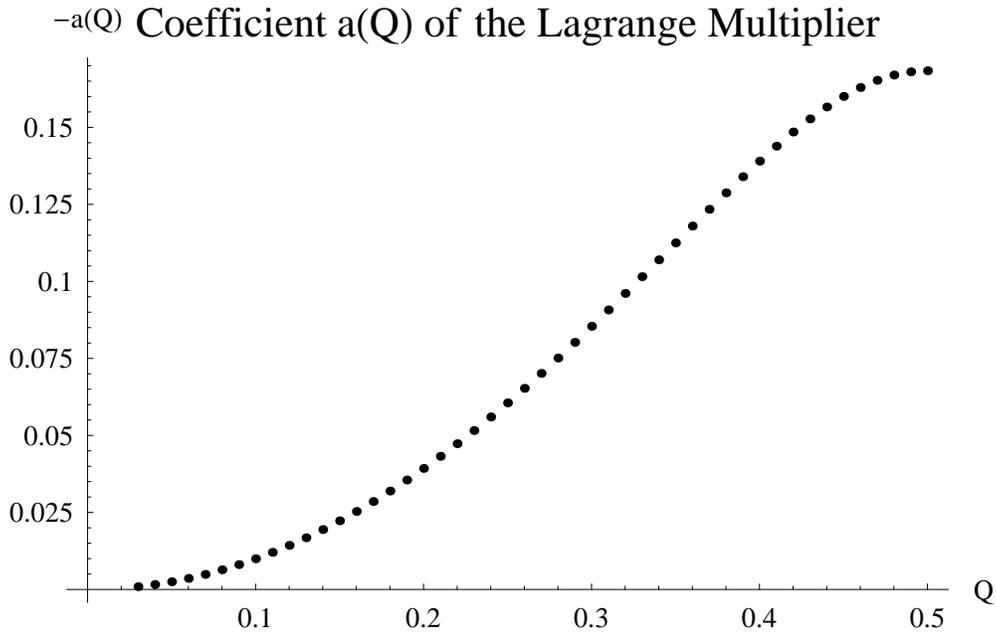}
\caption{Coefficient $a(Q)$ of the Lagrange multiplier $\langle i \lambda (x)\rangle$, see eq. (\ref{lambdaaQ}), in the $M= \infty$ generalization of the dual theory.}\label{aQ}
\end{center}
\end{figure}
\begin{figure}[t]
\begin{center}
\includegraphics[angle=-90,width = 0.8\textwidth]{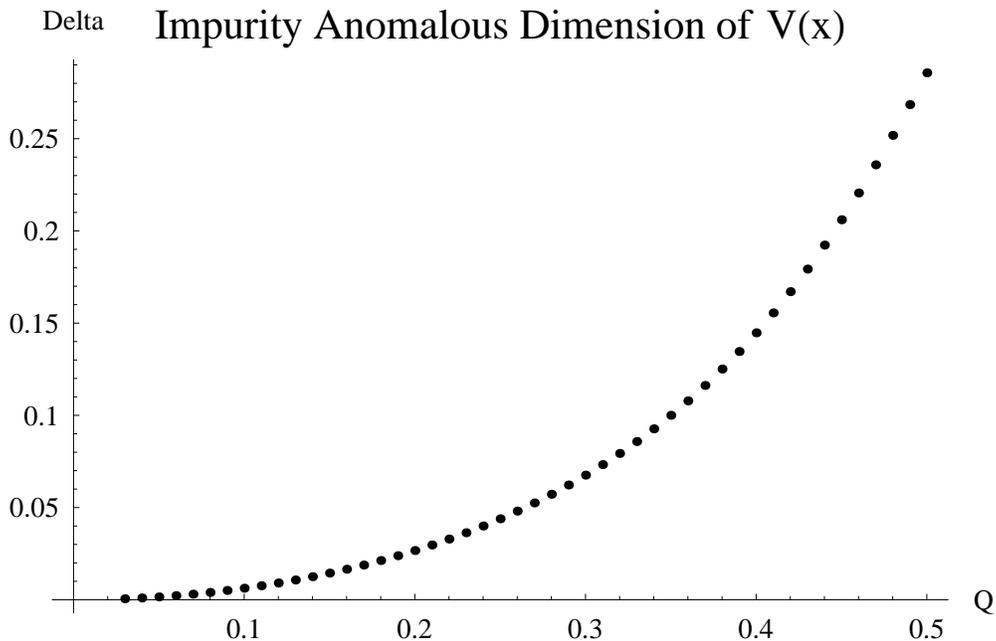}
\caption{Impurity anomalous dimension $\Delta^V_{\mathrm{imp}}$ of the monopole operator $V(x)$, see Eq.~(\ref{OPEfree}), computed in the $M = \infty$ generalization of the dual theory.} \label{FigDeltaN}
\end{center}
\end{figure}
\begin{figure}[t]
\begin{center}
\includegraphics[angle=-90,width = 0.8\textwidth]{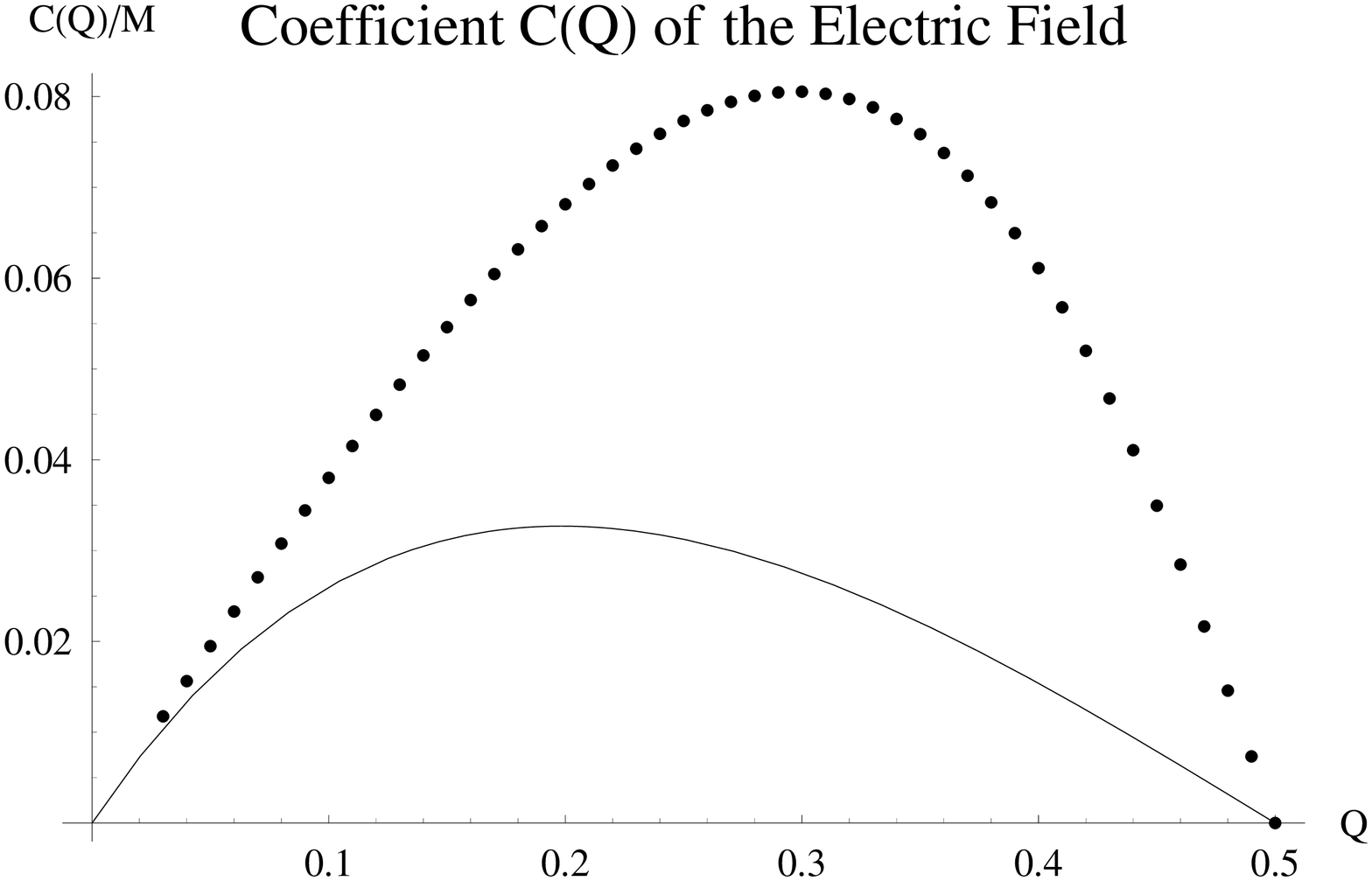}
\caption{Coefficient $C(Q)$ of the electric field, see eq. (\ref{EfieldQ}). The doted and solid curves corresponds to the strongly interacting theory at $M = \infty$ and the free theory respectively.} \label{CQ}
\end{center}
\end{figure}
The numerical solution for $a(Q)$ is shown in Fig. \ref{aQ}. We note that this solution agrees with the one obtained using the spectral form of propagator (\ref{DSpecN}). 

One can also attempt to use eq. (\ref{gapQa}) to find a series solution for $a(Q)$ near $Q = 0$. It is easy to convince oneself that, 
\beq \label{aQzero} a(Q) \approx -Q^2, \quad Q \to 0\eeq
Unfortunately, the integrand in eq. (\ref{gapQa}) is quite singular at $\nu \to 0$ for $a \to 0, Q \to 0$, so that a systematic series expansion beyond the leading order is not straight-forward. Nevertheless, we believe that such an expansion exists and $a(Q)$ is an analytic function of $Q$ near $Q = 0$. Assuming such analyticity and using charge conjugation symmetry, $a(Q) = a(-Q)$, one obtains, $a(Q) \approx -Q^2 + c_4 Q^4$ for $Q \to 0$. Here $c_4$ is a positive constant as the integral (\ref{gapQa}) diverges for $a < -Q^2$. 

Having found $a(Q)$ we immediately obtain the impurity anomalous dimension of the operator $V$ (given by eq. (\ref{DeltaVN1M})), see Fig. \ref{FigDeltaN}. This anomalous dimension is no-longer the trivial value $\Delta^V_{\mathrm{imp}} = |Q|$ of the free theory (\ref{DeltaVFree}). Given the leading behaviour of $a(Q)$ as $Q \to 0$ (\ref{aQzero}) and assuming analyticity of $a(Q)$ we conclude that $\Delta^V_{\mathrm{imp}}$ will also be analytic at $Q = 0$ (as opposed to the situation in the free theory). Moreover, 
\beq \label{DeltaVassym} \Delta^V_{\mathrm{imp}} \approx \sqrt{c_4}\, Q^2, \quad Q \to 0\eeq
 
Finally, we can now compute the coefficient of the electric field $C(Q)$. For the $M$-field generalization of the dual theory, we define the magnetic field by the same equation (\ref{diffprop}) as for $M=1$ theory, that is we consider the current associated with the global $U(1)$ symmetry,
\beq \label{diffpropN}
\langle -i F_{\mu}(x) \rangle = \langle -2 \pi i V^{\dagger}_{\alpha}\overleftrightarrow{\d_{\mu}}V_\alpha(x)\rangle=-2 \pi i \lim_{x \to y}(\d^x_{\mu}-\d^y_{\mu})\langle V_\alpha(x) V_\alpha^{\dagger}(y)\rangle = -2 \pi i M \lim_{x \to y}(\d^x_{\mu}-\d^y_{\mu}) D(x,y)\eeq
Due to our normalization of the $U(1)$ current, the electric field induced will be of order $M$. Now, differentiating $D(r,\theta)$ in (\ref{DintN}) and taking the symmetric limit as $\theta \to 0$,
\beq \label{CQN}-i \d_{\theta}D(r,0) = -\frac{1}{4 \pi r} \int_0^{\infty} d\nu \nu \tanh(\pi \nu) \frac{\sin(2 \pi 
Q)}{\cosh(2 \pi \sqrt{\nu^2+a}) - \cos(2 \pi Q)}\eeq
Using the values of $a(Q)$ found earlier (Fig. \ref{aQ}) and evaluating the integral (\ref{CQN}) numerically we obtain the coefficient $C(Q)$, shown in Fig. \ref{CQ} (dotted curve). Fig. \ref{CQ} also shows the value of $C(Q)$ in the free theory (\ref{CQfree}) for comparison (solid line). 

Alternatively, we can use (\ref{CQN}) to expand $C(Q)$ in a series in $Q$. Using the leading behaviour (\ref{aQzero}), we find,
\beq \label{CQassymN} C(Q) \approx M \left( \frac{\pi Q}{8} + \mathcal{O}(Q^3)\right), \quad Q \to 0\eeq
We see that the leading term in (\ref{CQassymN}) agrees with the one, which would be obtained by perturbation theory in $Q$ in the large $M$ limit (\ref{AN}), (\ref{CAconn}). It is also interesting to compare eq. (\ref{CQassymN}) to assymptotic behaviour of $C(Q)$ in the free theory (\ref{CQfreeassym}). We see that the leading term $C(Q)/M \approx \pi Q/8$ in both cases is the same, however, the subleading terms are different. The first subleading term in the free theory is non-analytic $\sim |Q| Q$, as opposed to the strongly interacting theory's analytic $\mathcal{O}(Q^3)$. Thus, we have been able to verify that the leading non-analyticity of $C(Q)$ in the free theory disappears in the interacting theory. We actually expect that the interacting theory cures itself of non-analyticities in $Q$ at all orders in $Q$.

Finally, let us discuss impurity anomalous dimensions of higher flux operators $V^q(x)$ for rational $Q = p/q$, as these are actual physical observables on the QED side of the duality. Once we go from $M = 1$ dual theory to its large $M$ counterpart, there are many possible generalizations of the $V^q(x)$ operator. Indeed, we can form different $SU(M)$ multiplets out of $q$ instances of $SU(M)$ fundamental $V_{\alpha}(x)$. We expect that these multiplets will have different (impurity) anomalous dimensions for $M$ finite. However, for $M = \infty$ all of these operators will have degenerate (impurity) anomalous dimensions. We can consider, for instance, the completely symmetric representation $V^q_{S}(x) = (V_\alpha(x))^q$, where $\alpha$ is some fixed index (no summation over $\alpha$). Then, for $M = \infty$,
\beq \langle V^q_S(x) (V^q_S(y))^{\dagger} \rangle = q! (D(x-y))^q\eeq
Hence, just as in the free case, the operator $V^q_S(x)$ has the impurity OPE (\ref{OPEVq}) with the corresponding impurity anomalous dimension,
\beq \Delta^V_{\mathrm{imp}}(q) = q \Delta^V_{\mathrm{imp}}\eeq

\section{Easy Plane Theory for general $N$}\label{sec:EasyPlane}

We now turn to the general case of the model $\mathcal{S}+\mathcal{S}_{\rm imp}$ with
a global U(1)$^{N-1}$ symmetry. The results of the previous section with $N=1$ can be rapidly
generalized, and will lead to a quantitative result for the scaling dimension of the monopole/VBS 
operator $V$ near the impurity.

\subsection{Duality in the Easy Plane Theory}\label{sec:EasyPlaneDual}
In this section, we consider a theory with $N$ flavours of spinon fields $z_{\alpha}$ ($N$ does not necessarily have to be large),
\beq \label{QEDN} L = \frac{1}{2 e^2} F^2_{\mu} + |(\d_{\mu}-i A_{\mu})z_{\alpha}|^2 + U(z_{\alpha})\eeq
Here, $U$ is some potential with the global $U(1)^{N}$ symmetry under independent phase rotations of the $z_{\alpha}$ fields.
The singlet component of this symmetry is actually gauged by the field $A_{\mu}$,
\beq\label{loc} U(1):\, z_{\alpha} \to e^{i \theta(x)} z_{\alpha}, \quad A_{\mu} \to A_{\mu} + \d_{\mu} \theta\eeq
while the non-singlet components are true global symmetries of the theory,
\beq\label{glob} U(1)^{N-1}:\, z_{\alpha} \to e^{i \theta^a t^a_{\alpha}} z_{\alpha}\eeq
where $t^a$, $a = 1 .. N-1$ are the generators of the $U(1)^{N-1}$ symmetry satisfying, $\sum_{\alpha} t^a_{\alpha} = 0$.
We require $U$ to
have a symmetry under the permutation of labels of $z_{\alpha}$ fields.  We choose $U$ in such a fashion that in the ``condensed" phase of the theory, it favours non-zero expectation values of all components of the $z_{\alpha}$ field, so that the vacuum manifold of the theory is a torus, $(S^1)^N$ (here we temporarily forget that the singlet symmetry is gauged). For $N = 2$ the theory under consideration is believed to describe the phase transition in the easy-plane antiferromagnet. 

We would like to dualize the theory (\ref{QEDN}). Similar theories were dualized in Ref.~\onlinecite{balents1,mv,SachdevReview,Sudbo,SudboBabaev}, and here we will
present a related discussion. An exact duality on the lattice appears in the compainon paper\cite{MSQ}, but we can write down the form of the dual action from very general considerations. Let us first identify the dual degrees of freedom. We go to the condensed phase of the theory (\ref{QEDN}), where all $\langle z_{\alpha} \rangle \neq 0$. Then, we can have vortices in any component of the $z_{\alpha}$ field. Formally, the homotopy group, $\pi_1((S^1)^N) = {\mathbb Z}^N$. So, we have $N$ types of vortices, which become the degrees of freedom of the dual theory $V_{\alpha}$, $\alpha = 1..N$. 

These vortices are global, rather than local. Indeed, let's consider a vortex in the first component $z_1$,
\beq z_1(\vec{x}) \sim v e^{i \lambda(\vec{x})}, \quad z_{\alpha} \sim v,\, \alpha \neq 1, \quad |\vec{x}| \to \infty\eeq
where $\lambda(\vec{x})$ winds from $0$ to $2 \pi$ as one goes around a contour out at infinity surrounding the vortex. Then, this vortex corresponds to a space-time dependent transformation of the vacuum (\ref{loc}), (\ref{glob}), with, $\theta(\vec{x}) = \frac{1}{N} \lambda(\vec{x})$ and $\theta^a(\vec{x}) t^a = (1-1/N,-1/N,...-1/N) \lambda(\vec{x})$. Thus, our vortex possesses a winding both in the local and in the global symmetry group. The winding in the local $U(1)$ group will be canceled by the gauge field,
\beq A_{\mu}(x) = \d_{\mu} \theta(x) = \frac{1}{N} \d_{\mu} \lambda(x)\eeq
hence our global vortices carry a magnetic flux $\Phi = 2 \pi/N$.\cite{Babaevfrac} Therefore, under the flux symmetry (\ref{Htransf}), the fields $V_{\alpha}$ should transform as,
\beq \label{VNtransf} V_{\alpha}(x) \to e^{2 \pi i \alpha(x)/N} V_{\alpha}(x)\eeq 
This fact will be crucial for the analysis to follow. 

The winding in the global group will lead to a long-range Coulombic interaction between our vortices. We will need dynamical gauge fields in the dual theory to give rise to this interaction. However, if we have a unit winding in each component of the $z$ field, our vortex becomes completely local, and carries total flux $2 \pi$. We can think of such a local vortex as a composite of $N$ global vortices of different types. The creation operator for this flux-tube, therefore, will be,
\beq \label{VN} {\cal V}(x) = \prod_{\alpha} V_{\alpha}(x)\eeq
Since the local vortex carries flux $2 \pi$, we can also associate the operator (\ref{VN}) with the monopole operator of the direct theory. Indeed, given (\ref{VNtransf}), under the flux symmetry (\ref{Htransf}),
\beq {\cal V}(x) \to e^{2 \pi i \alpha(x)} {\cal V}(x) \eeq
which is the correct transformation law for the monopole operator (\ref{Vqtransf}). 

We expect local vortices to interact by short range forces. Therefore, the operator (\ref{VN}) should not be charged under the emergent gauge fields of the dual theory. 

We are now ready to write down the dual theory,
\beq \label{dualN} L = \frac{1}{2 \tilde{e}^2} \sum_i (F^{\alpha}_{\mu})^2 +  |(\d_{\mu} - i B^{\alpha}_{\mu} - \frac{2 \pi i}{N} H_{\mu})V_{\alpha}|^2 + \tilde{U}(V_{\alpha})\eeq
Here $B^\alpha_{\mu} = B^{a}_{\mu} t^a_{\alpha}$, $a = 1..N-1$, are emergent dual gauge fields, which couple to the non-singlet currents. $F^{\alpha} = \epsilon_{\mu \nu \lambda} \d_{\nu} B^{\alpha}_{\lambda}$ are the corresponding field strengths. The dual potential $\tilde{U}(V_{\alpha})$ is chosen to have the same properties as the direct potential $U$: it has a $U(1)^N$ symmetry under independent phase rotations of the fields $V_{\alpha}$ and a symmetry under permutation of labels of $V_{\alpha}$ fields. Moreover, it favours $\langle V_{\alpha} \rangle \neq 0$ for all $\alpha$ in the condensed phase of the dual theory. Thus, the theory (\ref{dualN}) has a local $U(1)^{N-1}$ symmetry,
\beq \label{U1Nloc} U(1)^{N-1}: \quad V_\alpha(x) \to e^{i \phi^a(x) t^a_{\alpha}} V_{\alpha}(x), \quad B^a_{\mu} \to B^a_{\mu} + \d_{\mu}\phi^a\eeq
as well as the global $U(1)$ flux symmetry of the direct theory (\ref{VNtransf}) (which we have promoted to a local symmetry by introducing a non-dynamical source field $H_{\mu}$). As required, the monopole operator (\ref{VN}) is invariant under the local $U(1)^{N-1}$ symmetry of the dual theory (\ref{U1Nloc}).

The theory (\ref{dualN}) also has a global $U(1)^{N-1}$ symmetry associated with conservation of fluxes of the $N-1$ emergent gauge fields. This topological symmetry can be identified with the Noether's symmetry (\ref{glob}) of the direct theory.

\subsection{Wilson Loops in the Easy Plane Theory}
Now, we would like to apply the duality discussed in the previous sections to study the properties of Wilson loops
in the $U(1)^{N-1}$ symmetric theory (\ref{QEDN}). Recall, that to represent Wilson loops we must use a source field $H_{\mu}$ given by (\ref{HWilson}). As discussed for the case of $N=1$ theory, the effect of such a source field on the dual action (\ref{dualN}) is to introduce a twisted boundary condition for the vortex fields,
\beq \label{bcN} V_{\alpha}(\theta=2 \pi) = e^{-2 \pi i Q/N} V_{\alpha}(\theta = 0)\eeq
where $Q$ is the charge of our Wilson line. The physical origin of the factor $1/N$ is the fractional charge $2 \pi/N$ of the vortex fields $V_{\alpha}$ under the flux symmetry. Thus, we come to the amazing conclusion that the universal physics in the planar model is periodic in the charge $Q$ of the Wilson line, with period $Q = N$. This is a generalization of the $Q = 1$ periodicity of single flavour QED discussed before. As explained in a companion paper\cite{MSQ}, we expect that this $Q \sim N$ periodicity is a feature of the easy plane theory and does not generalize to the case with the full $SU(N)$ invariance.

Now, we would like to discuss more quantitative features of Wilson loops in the planar model. In particular, we would like to find the impurity anomalous dimension of the monopole operator (\ref{OPEfree}) and the coefficient of the electric field (\ref{EfieldQ}) at the critical point of the theory. We note that as in the $N=1$ case, we can easily dualize the magnetic field by differentiating the dual action with respect to the source field $H_{\mu}$,
\beq \label{FN} \langle - i F_{\mu} \rangle = \frac{(-2 \pi i)}{N} \langle V^{\dagger}_{\alpha} \overleftrightarrow{D}_{\mu} V_{\alpha}\rangle\eeq
with $D_{\mu}V_{\alpha} = (\d_{\mu} - i B^{\alpha}_{\mu} - \frac{2 \pi i}{N} H_{\mu})V_{\alpha}$.

To find $\Delta^V_{\mathrm{imp}}$ and $C(Q)$, we follow the procedure established for the $N=1$ case in section \ref{sec:dualLargeN} and perform a large $M$ expansion of the dual theory (\ref{dualN}). Namely, we promote each field $V_{\alpha}$ to an $SU(M)$ multiplet, $V^i_{\alpha}$, $i = 1..M$. Moreover, we replace the soft potential $\tilde{U}(V_{\alpha})$ by a hard constraint, $\sum_{i} |V^i_{\alpha}|^2 = 1/g$, for each $\alpha = 1..N$. This constraint will be enforced by a set of $N$ Lagrange multipliers $\lambda_{\alpha}$. Thus, our Lagrangian becomes,
\beq \label{DualM} L = \sum_{\alpha,i}|(\d_{\mu} - i B^{\alpha}_{\mu} - \frac{2 \pi i}{N} H_{\mu})V^i_{\alpha}|^2 +  \sum_{\alpha, i} i \lambda_{\alpha} (|V^i_{\alpha}|^2 - \frac{1}{g})\eeq
In (\ref{DualM}) we have also dropped the kinetic term for the gauge fields, as near the critical point such operators will be irrelevant.  In addition to the $U(1)_{\Phi}$ global flux symmetry and the $U(1)^{N-1}$ local symmetry of the original $M=1$ action, the theory (\ref{DualM}) also has a $SU(M)^{N}$ global symmetry under independent $SU(M)$ rotations of the $N$ $M$-tuplets $V^i_{\alpha}$. We note that the various $SU(M)$ multiplets talk to each other only through the gauge fields $B^{\alpha}_{\mu}$. 

We would like to generalize the observables of the $M = 1$ theory to the large $M$ case. The magnetic field (\ref{FN}) is generalized trivially,
\beq \label{FNM} \langle - i F_{\mu} \rangle = \frac{(-2 \pi i)}{N} \langle (V^i_{\alpha})^{\dagger} \overleftrightarrow{D}_{\mu} V^i_{\alpha}\rangle\eeq
The monopole operator (\ref{VN}) on the other hand, now carries indices under the $SU(M)^N$ group,
\beq \label{VNM} {\cal V}(x)_{i_1..i_N} = \prod_{\alpha} V^{i_{\alpha}}_{\alpha}(x)\eeq

The insertion of the Wilson loop source $H_{\mu}$ is again equivalent to the twisted boundary condition (\ref{bcN}).

We now perform a large $M$ expansion of the theory (\ref{DualM}) with the twisted boundary condition (\ref{bcN}), keeping $N$ fixed. We will be only able to make computations for $M = \infty$. We are interested in the physics at the critical point. We expand the theory about the saddle point $B^{\alpha}_{\mu} = 0$ (this is a saddle point as the twisted boundary condition (\ref{bcN}) does not couple to the non-singlet sectors of the theory \cite{footnote4}). As usual, the fluctuations of these gauge fields about the saddle point will be suppressed by powers of $1/M$. Thus, at $M = \infty$, we are left with $N$ decoupled instances of the Lagrangian (\ref{LXYN}) that has been discussed at length for the case of $N = 1$ theory. The only difference is the replacement, $Q \to Q/N$ in the boundary condition (\ref{bc}). Hence, we conclude,
\beq \label{VVNM} \langle {\cal V}(x)_{i_1..i_N} {\cal V}^{\dagger}(x')_{j_1..j_N}\rangle \stackrel{M = \infty}{=} \prod_{\alpha} \langle V^{i_{\alpha}}_{\alpha}(x) (V^{j_{\alpha}}_{\alpha})^{\dagger}(x')\rangle = D(x,y,Q/N)^N \prod_{\alpha} \delta_{i_{\alpha} j_{\alpha}}\eeq
where $D(x,x',Q)$ is the propagator in the $N = 1$ theory (\ref{LXYN}) with the twisted boundary condition (\ref{bc}) at $M =\infty$. The asymptotic behaviour of this propagator for $r \ll r'$ is given in eq. (\ref{assymDN}). Thus, the asymptotic behaviour of the correlation function (\ref{VVNM}) for $r \ll r'$ is
\beq \langle {\cal V}(x)_{i_1..i_N} {\cal V}^{\dagger}(x')_{j_1..j_N}\rangle \approx \left(\frac{1}{4 \pi r'}\right)^N \left(\frac{r}{r'}\right)^{N \sqrt{(Q/N)^2+a(Q/N)}} e^{-i Q \theta} G(\tau/r')\prod_{\alpha} \delta_{i_{\alpha} j_{\alpha}}, \quad |Q/N| < 1/2\eeq
where $G$ is some (known) function. Hence, the monopole operator ${\cal V}(x)$ in the planar $N$ component theory has the impurity OPE,
\beq \label{OPEVNM} {\cal V}(\vec{x}, \tau) \sim |\vec{x}|^{\Delta^{\cal V}_{\mathrm{imp}}} e^{-i Q \theta} \, {\cal V}_{\mathrm{imp}}(\tau)\quad \mathrm{for}\,\, |\vec{x}| \to 0\eeq
with 
\beq \label{DeltaVNM} \Delta^{\cal V}_{\mathrm{imp}} = N \sqrt{(Q/N)^2+a(Q/N)} = N \Delta^V_{N=1}(Q/N), \quad |Q/N| < 1/2\eeq
where the monopole impurity anomalous dimenension $\Delta^V_{N=1}(Q)$ in the $N = 1$, $M = \infty$ theory is given by Fig. \ref{FigDeltaN}.

From OPE (\ref{OPEVNM}), we observe that for integer $Q$ the monopole operator is single valued under $\theta \to \theta + 2 \pi$, even though the dynamical fields of the theory $V_{\alpha}$ obey twisted boundary conditions (\ref{bcN}). We also note that formulas (\ref{OPEVNM}) and (\ref{DeltaVNM}) are correct only for $|Q/N| < 1/2$; for other values of $Q$ they should be extended by periodicity $Q \sim Q + N$. 

We can now take the $N \to \infty$, $Q$-fixed limit of (\ref{DeltaVNM}). Using the assymptotic behaviour (\ref{DeltaVassym}), $\Delta^{\cal V}_{\mathrm{imp}} \sim Q^2/N$. Thus, the impurity anomalous dimension of the monopole operator is of order $\mathcal{O}(1/N)$ for $N \to \infty$ in the easy plane theory. It is interesting to note that, as discussed in section \ref{sec:CPN}, this is also true of the theory with a full $SU(N)$ symmetry. At this point, it is not clear whether this is just a coincidence.

Finally, let us discuss the special point $Q/N = 1/2$. Our interest in this point is not purely academic, as we expect $N = 2$, $Q = 1$ to correspond to the physical case of a single impurity in an easy plane antiferromagnet or superfluid. We recall that at this point the propagator $D(r,r',\theta,\tau)$  for $r \ll r'$ is dominated by two angular momenta, $l = \pm 1/2$,
\beq D(r,r',\theta,\tau) \approx \frac{1}{4 \pi r'} \left(\frac{r}{r'}\right)^{\sqrt{1/4+a(1/2)}} (e^{i \theta/2} + e^{-i \theta/2})B_{\sqrt{1/4+a(1/2)}}(\frac{\tau}{r'})\eeq
So that
\beq D(r,r',\theta,\tau)^N \approx \left(\frac{1}{4 \pi r'}\right)^N \left(\frac{r}{r'}\right)^{N \sqrt{1/4+a(1/2)}} \sum_{m=0}^{2 Q} \binom{2 Q}{m}e^{i(m-Q) \theta}\, G(\tau/r')\eeq
Hence, using (\ref{VVNM}), the correlation function of two monopole operators is dominated by angular momenta $l = -Q, -Q+1 .. Q-1, Q$ for $r \ll r'$. So, we conjecture the operator product expansion,
\beq \label{OPEVNMdeg} {\cal V}(\vec{x}, \tau) \sim \sum_{l=-Q}^{Q} c_l 
|\vec{x}|^{\Delta^{\cal V}_{l}} e^{-i l \theta} \, {\cal V}_{l}(\tau)\quad \mathrm{for}\,\, |\vec{x}| \to 0\eeq
At $M = \infty$ all the operators ${\cal V}_{l}$ have degenerate impurity anomalous dimensions $\Delta^{\cal V}_{l}$. As discussed in section \ref{sec:FreeTheory}, the anomalous dimensions of operators with oposite angular momenta are equal by CP symmetry emergent at the $Q/N = 1/2$ point. However, there is no fundamental reason why anomalous dimensions of operators with different values of $l$ should be equal. Thus, we expect the degeneracy to be lifted at higher orders in $1/M$ expansion. Therefore, unfortunately, the question of whether the OPE (\ref{OPEVNMdeg}) will be dominated by $l = 0$ or by finite $l$ is beyond the reach of our calculation. Nevertheless, our calculation at $M = \infty$  predicts for the physically relevant case of $N  = 2$, $Q = 1$,
\beq \Delta^{\cal V}_{\mathrm{imp}} \approx 0.57, \quad N = 2, Q = 1\eeq
The emergent $CP$ symmetry at the point $Q/N = 1/2$ means that quantum fluctuations manage to render the states of Figs. \ref{fig:vortex} and Fig. \ref{fig:antivortex} degenerate in the long-wavelength limit. We remind the reader the $CP$ symmetry is due to the emergent $Q \sim N$ periodicity of the easy plane theory.  No such periodicity is expected to occur in the full $SU(N)$ symmetric theory, where the impurity OPE is dominated by a single operator with a definite angular momentum as in eq. (\ref{vimp5}). 

For completeness sake, we also discuss the coefficient $C(Q)$ of the electric field. From eq. (\ref{FNM}) at $M = \infty$ we obtain,
\beq C(Q) = C_{N=1}(Q/N) \eeq
where the coefficient $C_{N=1}(Q)$ in the $N = 1$, $M = \infty$ theory is given by Fig. \ref{CQ}. We note that for $Q/N = 1/2$ the electric field vanishes, as it should, by the emergent CP symmetry.

\section{Conclusions}

This paper began with the theory $\mathcal{S}$ in Eq.~(\ref{cpn}) for square lattice quantum antiferromagnets in the
vicinity of a N\'eel-VBS quantum phase transitions. We considered generic local deformations of the antiferromagnet,
and argued that they could be classified into two categories. The first category, illustrated in Fig.~\ref{fig:bond}, is a modulated exchange impurity: we found an enhancement of VBS order, characterized by the exponent in Eq.~\ref{vimp3}. The second category
was realized by a missing or additional spin ({\em e.g.\/} Zn or Ni impurities on Cu sites), shown in Fig.~\ref{fig:vortex}.
For this case we found that VBS order was suppressed by the appearance of a VBS pinwheel, as in Fig.~\ref{fig:vortex},
and characterized by the scaling properties discussed in Section~\ref{sec:berry}.

The results of this paper should be useful in numerical studies of the quantum phase transition
between the N\'eel and VBS state \cite{sandvikdcp,melkokaul}. By enhancing an exchange constant as in Fig.~\ref{fig:bond},
and measuring the decay of the average VBS order parameter away from the impurity, the exponent $\Delta^V$
can be estimated from Eqs.~(\ref{vimp1}-\ref{vimp3}). There will be no mean VBS order in the vicinity of a missing
spin impurity as in Fig.~\ref{fig:vortex}. However, the spatial dependence in the VBS susceptibility is fixed by $\Delta^V_{\rm imp}$ in Eq.
(\ref{vimp5}). The positive value of $\Delta^V_{\rm imp}$ indicates that the VBS susceptibility should be suppressed near such
an impurity.

In STM studies of the cuprates, we have noted earlier the  
demonstration of bond-centered
charge order in the local density of states by Kohsaka {\em et al.} 
\cite{kohsaka}.
A numerical analysis of the pinning of such charge order by modulated  
exchange impurities
(in the class in Section~\ref{sec:exch}) has also been carried
out \cite{adrian,kivelson}. However, it is also experimentally  
possible to induce
``missing spin" impurities (in the class of Section~\ref{sec:berry})  
by replacing the Cu sites
with Zn and Ni impurities. There have been STM studies of such  
impurities \cite{yazdani,pan1,pan2},
and it be of great interest to carefully examine the nature
of the bond-centered modulations in the vicinity of such impurities.  
If we assume that the ``stripe" instability
is primarily associated with the appearance of magnetic order \cite 
{Zaanen89,Machida,Inui91,Kivelson03}, then the theory
of the enhancement of magnetic order near such impurities \cite 
{kolezhuk,MS} should apply: we should therefore
expect an increase in the strength of the density of states  
modulations in this model. In contrast, if we assume a VBS theory
of the modulations, then in the impurity model of Section~\ref 
{sec:berry}, the bond-centered modulations should be suppressed.
The experimental situation could well include both effects,  
complicating the interpretation.
However, evidence for VBS pinwheel configurations like
those in Fig.~\ref{fig:vortex} would lend  
strong support to the VBS theory.

\acknowledgments
We thank T.~Senthil, M.~Vojta and A.~Zhitnitsky for useful discussion.
This work was supported by NSF Grant No.\ DMR-0537077.

\appendix

\section{Electric Field in the Free Theory}\label{AppFieldFree}
We wish to use eq. (\ref{FieldProp}) to compute the electric field in the free theory. We start from the equation for the propagator (\ref{D2plus1}) and promote the $\omega$ integral to run over $D-2$ dimensions, as discussed in section \ref{sec:FreeTheory}, obtaining
\bea D(r,\theta) &=& \frac{\Gamma(2-D/2)}{(4 \pi)^{(D-2)/2}} \sum_l \frac{e^{i l \theta}}{2 \pi}\int du \,u^{D-3} J_{|l|}(u r)^2 \\&=&  \frac{\Gamma((3-D)/2)}{(4 \pi)^{(D-1)/2}} \frac{1}{r^{D-2}} \sum_l \frac{e^{i l \theta}}{2 \pi} \frac{\Gamma(|l|+D/2-1)}{\Gamma(|l|-D/2+2)}\eea
We see that the prefactor diverges for $D = 3$.  However, at $D = 3$ the sum over angular momenta becomes $\sum_l \frac{e^{i l \theta}}{2\pi} = \delta(\theta)$. So, we have to first perform the sum over angular momenta and then take the limit $D \to 3$. The sum over angular momenta can be performed in terms of hypergeometric functions, giving for $0 < Q < 1$,
\bea \label{Dhyperg}D(r,\theta) &=& \frac{\Gamma((3-D)/2)}{2^D \pi^{(D+1)/2}} \frac{e^{-i Q \theta}}{r^{D-2}} \Big(e^{i \theta} \frac{\Gamma(D/2-Q)}{\Gamma(3-D/2-Q)} F(\{1,D/2-Q\},\{3-D/2-Q\},e^{i \theta}) \nn \\&+&  \frac{\Gamma(D/2-1+Q)}{\Gamma(2-D/2+Q)} F(\{1,-1+D/2+Q\},\{2-D/2+Q\},e^{-i \theta})\Big)\nn\\\eea
where $F$ denotes the Barnes extended hypergeometric function. One can check that for $D=3$ the expression in brackets in (\ref{Dhyperg}) vanishes, cancelling the pole in the prefactor. Now, differentiating with respect to $\theta$,
\bea \label{diffDhyperg}-i \d_{\theta}D(r,\theta) &=& \frac{\Gamma((3-D)/2)}{2^D \pi^{(D+1)/2}} \frac{e^{-i Q \theta}}{r^{D-2}} \Big( \frac{(1-Q) \Gamma(D/2-Q)}{\Gamma(3-D/2-Q)} e^{i \theta} F(\{1,D/2-Q\},\{3-D/2-Q\},e^{i \theta})\nn\\
&+& \frac{\Gamma(D/2+1-Q)}{\Gamma(4-D/2-Q)} e^{2 i \theta} F(\{2,D/2-Q+1\},\{4-D/2-Q\},e^{i \theta})\nn\\
&-&\frac{Q \Gamma(D/2-1+Q)}{\Gamma(2-D/2+Q)}F(\{1,D/2-1+Q\},\{2-D/2+Q\},e^{-i \theta})\nn\\
&-& \frac{\Gamma(D/2+Q)}{\Gamma(3-D/2+Q)}e^{-i \theta} F(\{2,D/2+Q\},\{3-D/2+Q\},e^{-i \theta})\Big)\eea
According to (\ref{FieldProp}), to compute the electric field we need to take the limit as $\theta \to 0$ of (\ref{diffDhyperg}). Strictly speaking this limit does not exist as the hypergeometric functions blow up as $\theta \to 0$ (that is when the last argument goes to 1). However, we note that only the imaginary part of (\ref{diffDhyperg}) becomes infinite as $\theta \to 0$, while the real part has a well-defined limit. The expectation value of electric field $\langle - i E_r \rangle = - \langle - i F_{\theta} \rangle$ should  be real. Thus, we can drop the infinite imaginary part. Moreover, the imaginary part is antisymetric under $\theta \to - \theta$, so the ``symmetrized" limit of (\ref{diffDhyperg}) exists. It turns out that this symmetrized limit can be obtained by the formal summation formulas,
\bea \label{Fsymb}F(\{1,a\},\{b\},1) &=& \frac{1-b}{a-b+1}\\
F(\{2,a\},\{b\},1) &=& \frac{(b-1) (b-2)}{(a-b+1)(a-b+2)}\eea
So, taking $\theta \to 0$, plugging (\ref{Fsymb}) into (\ref{diffDhyperg}) and performing a few manipulations,
\beq -i \d_{\theta}D(\theta=0,r) = \frac{(2 Q-1) \Gamma((1-D)/2)}{2^{D+2}\pi^{(D+1)/2}} \frac{\Gamma(D/2+Q-1)}{\Gamma(1-D/2+Q)} \left(\frac{\sin(\pi (D/2+Q))}{\sin(\pi(D/2-Q))}-1\right)\frac{1}{r^{D-2}}\eeq
Taking the limit $D \to 3$,
\beq \label{diffDzer}-i \d_{\theta}D(\theta=0,r) = -\frac{1}{32 \pi r} (2 Q-1)^2 \tan (\pi Q)\eeq
Finally, plugging into (\ref{FieldProp}) we recover (\ref{EfieldQ}) with
\beq C(Q) = \frac{1}{8}(1-2 Q)^2 \tan (\pi Q), \quad 0 < Q <1 \eeq
We remind the reader that all the manipulations above have been performed for $0 < Q < 1$. The function $C(Q)$ can then be extended to other values of $Q$ by periodicity. In particular,  extending to the range $|Q| < 1$,
\beq \label{CQfreeApp} C(Q) = \frac{1}{8} (1-2|Q|)^2 \tan(\pi Q), \quad |Q| < 1\eeq

\section{Integral Form of the Twisted Propagator}\label{AppLinetProp}
In this section we review the derivation of the integral form of the twisted propagator (\ref{DFreeInt}) given in Ref. \onlinecite{Linet}. We use this integral form to compute the electric field (\ref{FieldProp}) and show that it is in agreement with the result obtained using spectral representation of the propagator (see Appendix \ref{AppFieldFree}). We also indicate how the free twisted propagator should be modified in the strongly interacting $M = \infty$ theory.

Recall the free massive propagator in $2D$ (without any twisted b.c.) obeys,
\beq \label{Prop2Deq} (-\d^2 + m^2) D(\vec{x},\vec{x}') = \delta(\vec{x}-\vec{x'})\eeq and is given by,
\beq \label{D2int} D_2(\vec{x},\vec{x}') = \frac{1}{2 \pi} K_0(m |\vec{x}-\vec{x}'|) = \frac{1}{2 \pi^2} \int_{-\infty}^{\infty} d\nu K_{i \nu}(m r) K_{i \nu}(m r') e^{\pi \nu} e^{-\nu |\theta-\theta'|}\eeq  where the integral representation is valid for $|\theta - \theta'| < 2 \pi$. The BesselK functions of imaginary argument satisfy the equation,
\beq \label{BesselKeq}\left(-\frac{1}{r}\frac{d}{dr}\left(r\frac{d}{dr}\right) -\frac{\nu^2}{r^2} + m^2\right) K_{i \nu}(m r) = 0\eeq Hence the functions $K_{i \nu}(m r) e^{\pm \nu \theta}$ are in the kernel of the operator $-\d^2_2 + m^2 = -\frac{1}{r}\frac{\d}{\d r}(r \frac{\d}{\d r}) - \frac{1}{r^2} \frac{{\d}^2}{\d \theta^2} + m^2$. Applying this operator to $D_2(\vec{x},\vec{x'})$ we learn,
\beq \label{BesselKcompl}\frac{1}{\pi^2 r^2} \int_{-\infty}^{\infty} d\nu \nu K_{i \nu} (m r) K_{i \nu} (m r') e^{\pi \nu} = \frac{1}{r} \delta(r-r')\eeq This identity will be useful to us later.

Now, we want to modify the propagator (\ref{D2int}) in such a way that it satisfies the twisted boundary conditions (\ref{bc}). Let's first symmetrize equation (\ref{D2int}) with respect to $\nu$ by noting $K_{i \nu} = K_{-i \nu}$.
Then, \beq D_2(r,r', \theta-\theta') = \frac{1}{2 \pi^2} \int_{-\infty}^{\infty} d\nu K_{i \nu}(m r) K_{i \nu}(m r') \cosh(\nu (\pi - |\theta-\theta'|))\eeq Now, we can generalize,
\beq \label{D2Q} D_2(r,r',\theta,Q) = \frac{1}{2 \pi^2} \int_{-\infty}^{\infty} d\nu K_{i \nu}(m r) K_{i \nu}(m r') \sinh(\pi \nu) U_{\nu}(\theta)\eeq
where \beq \label{Utrial}U_{\nu}(\theta)= \frac{\cosh(\nu(\pi-|\theta|))}{\sinh(\pi \nu)} + c(\nu) e^{\nu \theta} - c(-\nu) e^{- \nu \theta}\eeq  $D_2(r,r',\theta,Q)$ still satisfies eq. (\ref{Prop2Deq}) since, as noted above, the functions $K_{i \nu}(m r) e^{\pm \nu \theta}$ are in the kernel of $-\d^2_2 + m^2$. It remains to find $c(\nu)$ such that the propagator (\ref{D2Q}) obeys boundary conditions (\ref{bc}). After a few manipulations one arrives at,
\beq \label{UfullApp}U_{\nu}(\theta) = \frac{e^{- 2 \pi i Q \mathrm{sgn}(\theta)} \sinh(\nu |\theta|) + \sinh(\nu (2\pi -|\theta|))}{\cosh(2 \pi \nu)-\cos(2 \pi Q)}\eeq Next, one uses the identity,
\beq \sinh(\pi \nu) K_{i \nu}(m r) K_{i \nu}(m r') = \frac{\pi}{2} \int_{\xi_2}^{\infty} du\, J_0\big(m (2 r r')^{\frac12}(\cosh(u) - \cosh\xi_2)^{\frac12}\big)\sin(\nu u)\eeq
where $\xi_2 > 0$ is defined by,
\beq \cosh\xi_2 = \frac{r^2 + {r'}^2}{2 r r'}\eeq
Substituting this into (\ref{D2Q}),
\beq \label{D2uv} D_2(r,r',\theta,Q) = \frac{1}{2 \pi} \int_{\xi_2}^{\infty} du\, J_0\big(m (2 r r')^{\frac12}(\cosh(u) - \cosh\xi_2)^{\frac12}\big) \int_0^{\infty} d \nu \,U_{\nu}(\theta) \sin(\nu u)\eeq
We are mostly interested in the propagator with $r = r'$,
\beq D_2(r=r',\theta,Q) = \frac{1}{2 \pi} \int_{0}^{\infty} du\, J_0\big(m r \sqrt{2}(\cosh u - 1)^{\frac12}\big) \int_0^{\infty} d \nu \,U_{\nu}(\theta) \sin(\nu u)\eeq In principle, it is possible to perform the integral over $\nu$ analytically in (\ref{D2uv}) (see Ref.~\onlinecite{Linet}). This, however, will not be very benificial for our purposes. Instead, let's proceed directly to the three-dimensional massless propagator, obtained by integrating over the mass parameter of the two dimensional propagator (\ref{D2plus1}),
\beq D(r,\theta) = \frac{1}{2 \pi^2 r \sqrt{2}} \int_0^{\infty} du \frac{1}{(\cosh u - 1)^{\frac12}}\int_0^{\infty}d \nu \, U_{\nu}(\theta) \sin(\nu u)\eeq
where we have computed only the 3 dimensional propagator with $r = r'$, $\tau = \tau'$. Now, performing the integral over $u$,
\beq D(r,\theta) = \frac{1}{4 \pi r} \int_0^{\infty} d\nu \tanh(\pi \nu) U_{\nu}(\theta)\eeq
To find the electric field we again use eq. (\ref{FieldProp}),
\bea -i \d_{\theta}D(r,\theta) &=& -\frac{1}{4 \pi r} \int_0^{\infty} d\nu\, \nu \tanh(\pi \nu) \Big(\frac{\sin(2 \pi Q) \cosh(\nu \theta)}{\cosh(2 \pi \nu) - \cos(2 \pi Q)} \nn\\&+& i \mathrm{sgn(\theta)} \frac{(\cos(2 \pi Q) \cosh(\nu \theta) - \cosh(\nu (2\pi - |\theta|)))}{\cosh(2 \pi \nu) - \cos(2 \pi Q)}\big)\eea
Again, the real part of $-i \d_{\theta}D(r,\theta)$ has a well-defined limit as $\theta \to 0$, while the imaginary part is antisymmetric under $\theta \to -\theta$ and diverges as $\theta \to 0$. So the ``symmetrized" limit is given by,
\beq -i \d_\theta D(r,\theta = 0) = -\frac{1}{4 \pi r} \int_0^{\infty} d\nu \, \nu \frac{\sin(2 \pi Q) \tanh(\pi \nu)}{\cosh(2 \pi \nu)-\cos(2 \pi Q)} = -\frac{1}{32 \pi r} (2 |Q|-1)^2 \tan(\pi Q)\eeq
in agreement with an earlier computation (\ref{diffDzer}) based on spectral decomposition. Thus, $C(Q)$ is again given by expression (\ref{CQfree}). 

Now we generalize the above derivation of the twisted propagator to the strongly interacting $M = \infty$ theory. The strongly interacting theory differs from the free theory by the aditional space-varying potential $\langle i \lambda(\vec{x},\tau)\rangle$, so that the propagator satisfies,
\beq \label{AppPropEqa} (-\d^2 + \frac{a(Q)}{|\vec{x}|^2}) D(x,x',Q) = \delta(x-x')\eeq
 We again rewrite $D(x,x',Q)$ in terms of the two dimensional massive propagator $D_2(\vec{x},\vec{x}',m^2, Q)$ as in eq. (\ref{D2plus1}). The two dimensional propagator satisfies,
\beq \label{D2Neq}\left(-\frac{1}{r} \frac{\d}{\d r}(r \frac{\d}{\d r}) -\frac{1}{r^2}\frac{\d^2}{\d\theta^2} + \frac{a}{r^2} + m^2\right)D_2(r,r',\theta,\theta') = \delta(\vec{x}-\vec{x}')\eeq
We need to generalize the two-dimensional, massive, twisted, free propagator (\ref{D2Q}) so that it obeys the above equation. We observe that the function $U_{\nu}(\theta)$ (\ref{Utrial}), (\ref{UfullApp}) satisfies,
\beq \label{Ueq}\frac{\d^2 U_{\nu}}{\d \theta^2} = \nu^2 U_{\nu}(\theta) -2 \nu \delta(\theta)\eeq
Now combining eqs. (\ref{BesselKeq}), (\ref{BesselKcompl}) and (\ref{Ueq}), we find that,
\beq \label{D2BesselN}D_2(r,r',\theta,Q) = \frac{1}{\pi^2}\int_0^{\infty} d\nu K_{i \nu} (m r) K_{i \nu}(m r') \sinh(\pi \nu) \frac{\nu}{\sqrt{\nu^2 + a}} U_{\sqrt{\nu^2 + a}}(\theta)\eeq
satisfies (\ref{D2Neq}) as needed. Proceeding as above from two to three dimensional propagator, and setting $r = r'$, $\tau = \tau'$
\beq \label{AppDintN} D(r,\theta) = \frac{1}{4 \pi r} \int_0^{\infty} d\nu \tanh(\pi \nu) \frac{\nu}{\sqrt{\nu^2+a}} U_{\sqrt{\nu^2 + a}}(\theta)\eeq


\begin{thebibliography}{10}
\bibitem{vbscharge} In an antiferromagnet of $S=1/2$ degrees of freedom,
VBS order is associated with modulations in the expectation values of 
the bond exchange energy $\vec{S}_1 \cdot \vec{S}_2$, where $\vec{S}_{1,2}$ 
are the spin operators on the ends of the bond. While this is the
best physical interpretation of VBS order, in a symmetry classification VBS order 
can also be considered to be a particular realization of ``charge order". This is because
in the underlying Hubbard-like model (with a respulsive energy $U$
and hopping matrix element $t$), from which the antiferromagnet descends, 
the VBS state will have modulations of order $t/U$ in the charge density in the
bonding orbital between the two sites. For a discussion of how the VBS characterization
of the charge order can explain the experimental observations in the cuprates,
see Ref.~\onlinecite{poilblanc,vojta}.
\bibitem{footnote1} Here we use the notation that $V^{q=1}(x) \sim V(x)$ and $V^{q=-1}(x) \sim V^{\dagger}(x)$. The precise proportionality factor between $V$'s in the direct and dual picture is a delicate matter (related in part to the precise definition of the monopole operator in the direct picture), which shall not be very important to us here.
\bibitem{footnote2} Given our normalization of $A$ one has to multiply the value of universal conductance presented in 
Refs.~\onlinecite{Fazio,FisherGirvin} by $2 \pi$.
\bibitem{footnote3} In $D = 3$ we take the $P$-parity symmetry to correspond to a reflection about one spatial axis, $\theta \to -\theta$ (a complete spatial inversion $\vec{x} \to -\vec{x}$ is a rotation in $2+1$ dimensions). The monopole field $V$ is a pseudoscalar, which means that under charge conjugation $C$, $V \to V^{\dagger}$ and under $P$-parity, $V(x_1,x_2,x_3) \to V^{\dagger}(x_1,-x_2,x_3)$.
\bibitem{footnote4} In reality, we expect a non-zero $B^{\alpha}_{\mu}$, corresponding to finite flavour charge density  of the direct theory. Indeed, we expect that the external charge will be screened in the direct theory by a flavoured spinon $z_{\alpha}$. The associated flavour charge density (magnetization) in the $SU(N)$ symmetric case has been discussed in Ref.~\onlinecite{MS}. However, at the critical point, the screening cloud will be distributed over the whole size of the system. Therefore, the corresponding (finite) flux density will be non-intensive and will not affect intensive observables such as $\Delta^V_{\mathrm{imp}}$ and $C(Q)$.
\end{thebibliography}
\end{document}